\documentclass[journal]{IEEEtran}
\usepackage[bookmarks,colorlinks]{hyperref}
\usepackage{enumitem}
\usepackage[linesnumbered,ruled,lined]{algorithm2e}

\usepackage{graphicx}
\usepackage{dcolumn}
\usepackage{bm}

\usepackage[usenames,dvipsnames]{xcolor}

\usepackage{tabularx}
\usepackage{array}

\usepackage{amsmath,amsfonts}

\usepackage{array}
\usepackage[caption=false,font=normalsize,labelfont=sf,textfont=sf]{subfig}
\usepackage{textcomp}
\usepackage{stfloats}
\usepackage{url}
\usepackage{verbatim}
\usepackage{graphicx}
\usepackage{cite}
\usepackage{svg}
\usepackage[noend]{algpseudocode} 
\usepackage{etoolbox}

\usepackage{float}
\usepackage{pgfplots}
\usepackage[bookmarks,colorlinks]{hyperref}
\usepackage[linesnumbered,ruled,lined]{algorithm2e}
\usepackage{enumitem}
\usepackage{algpseudocode}
\usetikzlibrary{shapes.multipart,intersections}
\usepackage{cite}
\usepackage{amsmath,amssymb,amsfonts,amsthm,steinmetz}
\usepackage{graphicx}
\usepackage{mathrsfs}  
\usepackage{textcomp}
\usepackage{acronym}
\usepackage{xcolor}
\usepackage{upgreek,xspace}

\usepackage{tikz}
\usetikzlibrary{calc}
\makeatletter
\newcommand{\gettikzxy}[3]{%
  \tikz@scan@one@point\pgfutil@firstofone#1\relax
  \edef#2{\the\pgf@x}%
  \edef#3{\the\pgf@y}%
}
\makeatother

\usepackage[draft]{todonotes}

\usepackage{esvect}

\usepackage{times}
\usepackage{bm}
\usepackage{amsmath}
\usepackage{amssymb}
\usepackage{stmaryrd}
\usepackage{babel}
\usepackage{graphics, graphicx}
\usepackage{xcolor}
\usepackage{gensymb}
\usepackage{cite}
\usepackage{enumitem}
\usepackage{url}

\ifCLASSINFOpdf
\else
\fi
\begin{document}
%
\title{Virtual VNA: \\Minimal-Ambiguity Scattering Matrix Estimation \\with a Fixed Set of ``Virtual'' Load-Tunable Ports}
%
%
%

\author{Philipp~del~Hougne,~\IEEEmembership{Member,~IEEE}
\thanks{The author acknowledges funding from the IETR PEPS program (project ``IMPEST''), the ANR France 2030 program (project ANR-22-PEFT-0005), the European Union's European Regional Development Fund, and the French region of Brittany and Rennes Métropole through the contrats de plan État-Région program (projects ``SOPHIE/STIC \& Ondes'' and ``CyMoCoD'').}
\thanks{P.~del~Hougne is with Univ Rennes, CNRS, IETR - UMR 6164, F-35000 Rennes, France (e-mail: philipp.del-hougne@univ-rennes.fr).}
}

\maketitle

\begin{abstract}
We estimate the scattering matrix of an \textit{arbitrarily complex} linear, passive, time-invariant system with $N$ monomodal lumped ports by inputting and outputting waves only via a fixed set of $N_\mathrm{A}<N$ ports while terminating the remaining $N_\mathrm{S}=N-N_\mathrm{A}$ ``not-directly-accessible'' (NDA) ports with tunable individual loads. First, we present a closed-form approach requiring at least three arbitrary, distinct, and known loads at each NDA port; sign ambiguities on off-diagonal scattering coefficients associated with NDA ports are inevitable. Being matrix-valued, our approach is ideally suited to mitigate noise sensitivity using more accessible ports. It also yields $1+2N_\mathrm{S}+N_S(N_S-1)/2$ as upper bound on the number of required measurements $N_\mathrm{cal}$ for $N_\mathrm{A}>1$ in the low-noise regime. Second, we present a gradient-descent approach using (potentially opportunistic) random load configurations, enabling flexible adjustments of $N_\mathrm{cal}$ to further mitigate noise. Third, we present an intensity-only gradient-descent approach that dispenses with phase-sensitive detectors at the expense of an additional blockwise phase ambiguity. Then, we discuss in what applications the inevitable remaining ambiguities are problematic and how they can be lifted. Finally, we experimentally validate all three approaches with an eight-port reverberation chamber and $N_\mathrm{A}=N_\mathrm{S}=4$, systematically assessing the sensitivity to noise and $N_\mathrm{cal}$. We coin our technique ``virtual vector network analyzer (VNA)'' because it implies that suitably tunable and characterized individual loads can essentially be interpreted as additional ``virtual'' (as opposed to actual) VNA ports. Our method can not only characterize static large antenna systems and circuits with many and/or embedded ports but also reconfigurable wave systems (smart radio environments, physical neural networks, programmable photonic integrated circuits); it may furthermore enable wireless sensing paradigms in indoor surveillance, non-destructive testing, and bioelectronics. 
\end{abstract}

\begin{IEEEkeywords}
Virtual VNA, tunable load, impedance matrix estimation, scattering matrix estimation, contactless antenna characterization, reconfigurable intelligent surface, end-to-end physics-compliant channel estimation, phase retrieval, ambiguity.
\end{IEEEkeywords}

\IEEEpeerreviewmaketitle

\section{Introduction}
\label{sec_introduction}

Any linear, passive, time-invariant wave system connected via $N$ monomodal ports to the outside world is fully characterized by its scattering matrix $\mathbf{S} \in \mathbb{C}^{N \times N}$ which relates the incoming fields $\mathbf{x} \in \mathbb{C}^{N\times1}$ and outgoing fields $\mathbf{y} \in \mathbb{C}^{N\times1}$: $\mathbf{y} = \mathbf{S} \mathbf{x}$. 
Measuring a system's scattering matrix is hence a common prerequisite to controlling the system's interactions with waves. In the microwave regime, standard laboratory equipment known as vector network analyzer (VNA) is capable of accurately measuring the entire scattering matrix by injecting and receiving waves via all system ports. 
However, in important scenarios on which we elaborate below, some of the system's ports are not directly accessible (NDA) to inject and receive waves: in some cases, certain ports are simply physically inaccessible; in other cases, the ports are so numerous that accessing all ports to inject and receive waves is effectively not possible for practical purposes. In both cases, a direct measurement of the system's  full scattering matrix with the conventional VNA method is thus not feasible.

A core question, both fundamentally and for applications, is hence whether it is possible to determine the system's full $N \times N$ scattering matrix by exciting the system, and observing the scattered fields, only via $N_\mathrm{A}<N$ accessible ports. Of course, the  scattering observable via these $N_\mathrm{A}$ ports depends on the boundary conditions imposed at the remaining $N_\mathrm{S} = N-N_\mathrm{A}$ NDA ports. In this paper, we explore the case in which these boundary conditions are tunable: although $N_\mathrm{S}$ ports are NDA to input/output waves, their terminations with load impedances are to some extent controllable. To be clear, this paper is concerned with scenarios involving a fixed set of accessible ports and a fixed set of NDA ports terminated with tunable loads -- as opposed to being able to reassign ports from one set to the other between measurements.

The raised question relates to various research efforts on multi-port network measurement techniques. When the system ports are physically accessible but their number $N$ exceeds the number $N_\mathrm{A}$ of VNA ports, the traditional measurement technique consists in several $N_\mathrm{A}$-port measurements connecting the VNA to different combinations of system ports while terminating the remaining system ports with auxiliary matched loads~\cite{tippet1982rigorous,ruttan2008multiport,2023paper}. This approach is prone to inaccuracies, arduous and not scalable to cases with large $N$.\footnote{To facilitate the countless required changes of connections, commercial vendors offer full cross-bar switch matrices. However, due to the required number of switches, these are very costly and also not scalable to cases with large $N$.} 
To reduce the number of required changes of connections between system ports and VNA ports, one can (for a given set of such connections) alternate between different loads terminating the system ports not connected to the VNA. This family of methods has been the topic of academic research for decades under terms like ``unterminating''~\cite{bauer1974embedding} (only studied for $N_\mathrm{S}=1$) and ``port reduction''~\cite{lu2000port,lu2003multiport,pfeiffer2005recursive}. The latter typically differs from the problem studied in this paper in that the sets of accessible ports are reassigned between different sets of measurements, which precludes an application to scenarios in which NDA ports are physically inaccessible, e.g., because they are embedded in a circuit or antenna system.\footnote{For the same reason, a research line specific to antenna array characterization that relies on measuring embedded element patterns~\cite{buck2022measuring} differs from the problem studied in the present paper.} 

The literature contains various experimental works addressing parts of the problem studied in the present paper (sometimes in special scenarios) for the case of $N_\mathrm{S}=1$~\cite{garbacz1964determination,bauer1974embedding,mayhan1994technique,davidovitz1995reconstruction,pfeiffer2005recursive,pfeiffer2005equivalent,pfeiffer2005characterization,pursula2008backscattering,bories2010small,van2020verification,sahin2021noncontact,kruglov2023contactless}, and two for the case of $N_\mathrm{S}=2$~\cite{wiesbeck1998wide,monsalve2013multiport}. Indeed, some of these works make specific assumptions about the system under test (preventing their applicability to arbitrarily complex systems). Moreover, these works only include closed-form scalar-valued methods that fundamentally struggle with measurement noise. To the best of our knowledge, generic matrix-valued closed-form methods have not been proposed, nor have gradient-descent methods been proposed, let alone for intensity-only measurements. We are also not aware of experimental validations of such techniques with $N_\mathrm{S}>2$ and/or on highly complex systems. We fill these research gaps in the present paper.

Furthermore, so far, our discussion (in line with existing literature) was focused on the characterization of \textit{static} systems. As far as we know, it has not been noticed that these methods are also applicable to the characterization of \textit{tunable} wave systems such as reconfigurable intelligent surfaces, optical neural networks or photonic integrated circuits. Indeed, tunable lumped elements in reconfigurable wave systems can be understood as virtual ports terminated by tunable loads. This insight remarkably extends the scope of relevance of the studied methods, and opens up new possibilities in wireless sensing (a brief discussion is provided in Sec.~\ref{sec_discuss_ambiguity}).

We coin the presented techniques ``Virtual VNA'' because they imply that any suitably tunable and characterized load or lumped element can essentially serve as an additional ``virtual'' (as opposed to actual) VNA port. The only caveat is that a few  sign ambiguities are inevitable. As discussed in Sec.~\ref{sec_discuss_ambiguity}, these ambiguities are unproblematic in some applications and can be lifted by various modifications of the problem statement in other applications.

Our key contributions are summarized as follows:
\begin{enumerate}
    \item We present a \textit{closed-form method} that only assumes linearity, passivity, time-invariance, reciprocity and monomodal lumped ports, and that is inherently matrix-valued, thus capable of optimally coping with measurements from arbitrarily many accessible ports. Our method identifies fundamental requirements: Three distinct and known loads must be available at each NDA port; specific characteristics of the three loads (such as emulating certain calibration standards) are not required, and the set of available loads can be different at each NDA port. Our method also identifies inevitable remaining ambiguities on signs of off-diagonal scattering coefficients associated with NDA ports. Moreover, our method implies an upper bound on the number of required measurements: $1+2N_\mathrm{S}+N_S(N_S-1)/2$ if $N_\mathrm{A}>1$, else $N_S(N_S-1)/2$ additional measurements are needed. 
    \item We present an alternative \textit{gradient-descent method} which is compatible with (potentially opportunistic) random load configurations and can flexibly incorporate additional measurements, e.g., to mitigate noise. Moreover, the number of NDA ports whose terminations differ between subsequent measurements typically strongly exceeds one or two, which also improves the noise robustness. In the low-noise regime, the number of required measurements for a desired accuracy can remain below the above-mentioned upper bound. 
    \item We adapt the gradient-descent method to work purely based on \textit{intensity-only measurements}, which entails an additional blockwise phase ambiguity but drastically alleviates the detection hardware requirements.
    \item We discuss whether the inevitable remaining ambiguities are problematic in various applications in wireless sensing and system characterization, and we identify various opportunities to lift the ambiguities.
    \item We experimentally validate all methods (with $N_\mathrm{A}=N_\mathrm{S}=4$) in a highly complex eight-port system  based on a reverberation chamber for which clearly no a priori knowledge is available. We systematically study the sensitivity to noise and the number of measurements.
\end{enumerate}

\textit{Outline:} In Sec.~\ref{sec_principle}, we describe the principle of the  ``Virtual VNA'' concept and explain how it relates to other topics of contemporary research interest.
In Sec.~\ref{Sec_theory_methods}, we present the theory and methodology of our three approaches: a matrix-valued closed-form approach (Sec.~\ref{subsec_CLFO}), a gradient-descent approach (Sec.~\ref{subsec_gradesc}), and an intensity-only gradient-descent approach (Sec.~\ref{subsec_phaseless}).
In Sec.~\ref{sec_discuss_ambiguity}, we discuss whether remaining inevitable ambiguities are problematic and how they can be lifted.
In Sec.~\ref{sec_exp}, we describe the experimental validation of our three approaches.
In Sec.~\ref{sec_discussion}, we comprehensively compare the ``Virtual VNA'' to existing related techniques and concretely describe application perspectives.
We close with a conclusion in Sec.~\ref{sec_conclusion}.

\textit{Notation:} The superscripts $^T$ and $^\dagger$  denote the transpose and the transpose conjugate, respectively. $\mathbf{A}_\mathcal{BC}$ denotes the block of the matrix $\mathbf{A}$ comprising rows [columns] whose indices are in the set $\mathcal{B}$ [$\mathcal{C}$]. $\mathbf{A}_{\mathcal{B}_i\mathcal{C}_j}$ denotes the entry of the matrix $\mathbf{A}$ whose row [column] index is the $i$th [$j$th] element of the set $\mathcal{B}$ [$\mathcal{C}$].
$\mathrm{Tr}(\mathbf{A})$ denotes the trace of the matrix $\mathbf{A}$.
$\mathbf{I}_d$ denotes the $d \times d$ identity matrix.
$\jmath = \sqrt{-1}$ denotes the imaginary unit.

\section{Principle}
\label{sec_principle}

\subsection{Problem Statement}
\label{subsec_ProblemStatement}

\begin{figure}[t]
    \centering
    \includegraphics[width=0.9\columnwidth]{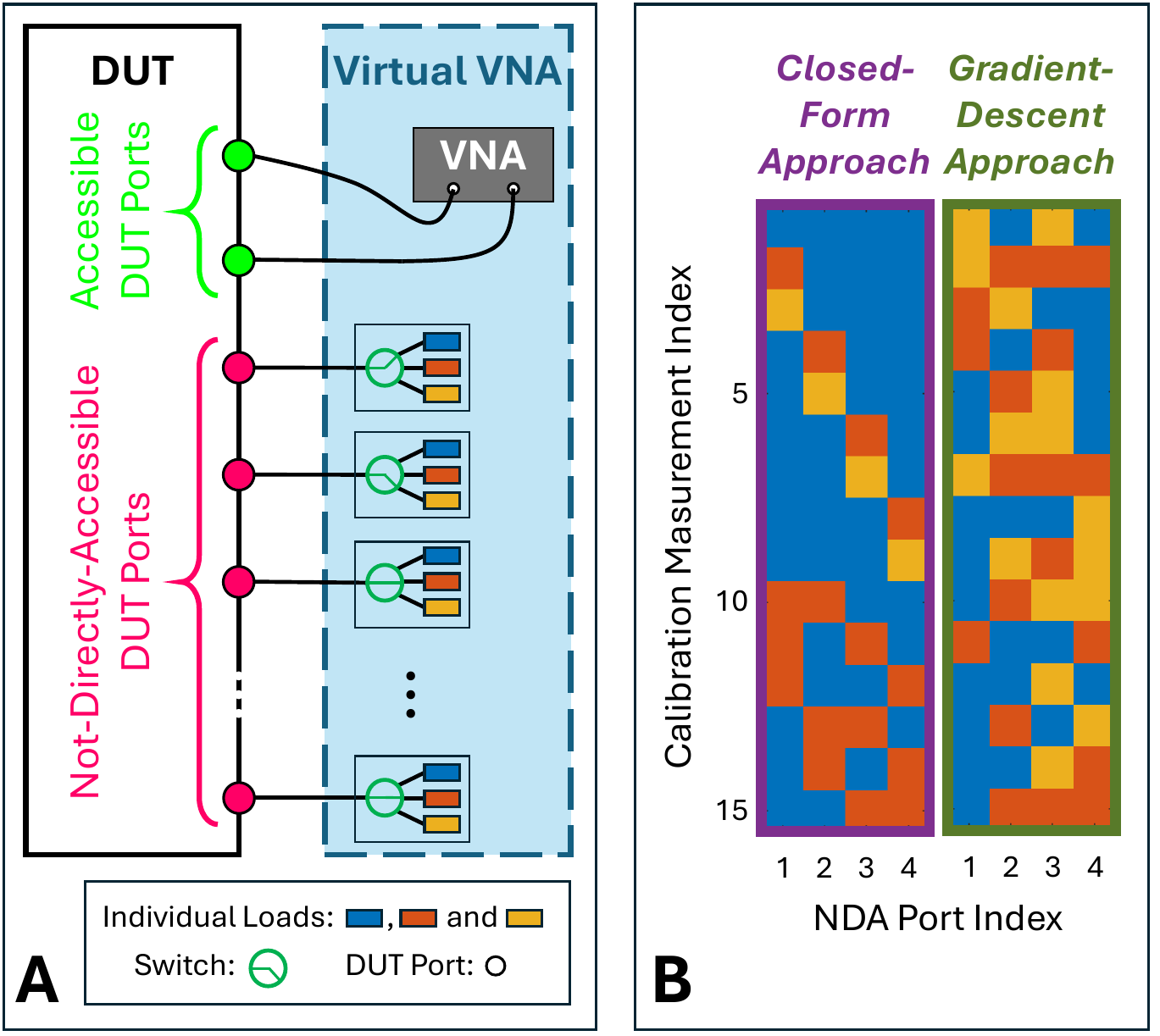}
    \caption{``Virtual VNA'' principle. (A) The monomodal lumped ports of a DUT are split into ``accessible'' (light green) and ``not-directly-accessible'' (NDA, pink) ones. We mean by NDA that waves cannot be injected and received (irrespective of whether physical access is in principle possible), but that other manipulations such as load switching are possible. The DUT's accessible ports are connected to a conventional VNA. The DUT's NDA ports are connected to individual tunable loads. The ensemble of the conventional VNA and the tunable individual loads, in combination with the data analysis developed in this paper, is referred to as ``Virtual VNA'' because each tunable individual load plays the role of a virtual (rather than actual) VNA port. (B) Illustration of the series of required load configurations for the closed-form approach (purple frame, left, Algorithm~\ref{Alg1}) and the gradient-descent approaches (green frame, right, example for $N_\mathrm{cal}=15$, Algorithm~\ref{Alg2} and Algorithm~\ref{Alg3}) for a scenario with $N_\mathrm{S}=4$ NDA ports.}
    \label{Fig0}
\end{figure}

Our goal is to estimate the $N\times N$ scattering (or impedance) matrix of a device under test (DUT) with $N$ lumped monomodal ports. We assume that a subset of $N_\mathrm{A}<N$ DUT ports can be directly connected to an $N_\mathrm{A}$-port VNA; we refer to these ports as being ``accessible'' to inject and receive waves. The remaining $N_\mathrm{S}=N-N_\mathrm{A}$ ports are ``not-directly-accessible'' (NDA), meaning that we cannot inject or receive waves via them (and hence not connect them to VNA ports); however, we assume that each NDA port can be terminated with three different and known loads. A schematic illustration of this problem is provided in Fig.~\ref{Fig0}. 

By terminating the $N_\mathrm{S}$ NDA ports with different sets of known load configurations (illustrated in Fig.~\ref{Fig0}B) and measuring the $N_\mathrm{A}\times N_\mathrm{A}$ scattering (or impedance) matrix at the DUT's accessible port for each configuration, we seek to estimate the DUT's $N\times N$ scattering (or impedance) matrix. 
Importantly, whether a port is accessible or NDA remains fixed throughout all measurements.

To formalize the problem, we denote by $\mathcal{A}$ and $\mathcal{S}$ the sets of indices associated with accessible and NDA ports, respectively; the cardinality (number of elements) of these sets is $|\mathcal{A}| = N_\mathrm{A}$ and  $|\mathcal{S}| = N_\mathrm{S}$. Given our separation of the ports into accessible and NDA ones, a $2\times 2$ partition of $\mathbf{S}$ and $\mathbf{Z}$ naturally ensues:\footnote{Scattering and impedance matrices contain the same information and are related via $\mathbf{Z} = Z_0 (\mathbf{I}_N + \mathbf{S}) (\mathbf{I}_N - \mathbf{S})^{-1}$, where $Z_0$ is the characteristic impedance of the asymptotic scattering channels (e.g., single-mode transmission lines such as coaxial cables) connected to the ports.}
\begin{equation}
    \mathbf{S} = \begin{bmatrix} \mathbf{S}_\mathcal{AA} & \mathbf{S}_\mathcal{AS}\\ \mathbf{S}_\mathcal{SA} & \mathbf{S}_\mathcal{SS}   \end{bmatrix} \ \ ; \ \   \mathbf{Z} = \begin{bmatrix} \mathbf{Z}_\mathcal{AA} & \mathbf{Z}_\mathcal{AS}\\ \mathbf{Z}_\mathcal{SA} & \mathbf{Z}_\mathcal{SS}   \end{bmatrix}.
    \label{eq1}
\end{equation}
For conciseness, we do not explicitly print the frequency dependence in Eq.~(\ref{eq1}) and the following.
The scattering matrix $\hat{\mathbf{S}} \in \mathbb{C}^{N_\mathrm{A}\times N_\mathrm{A}}$ that can be measured at the $N_\mathrm{A}$ accessible ports depends on the loads terminating the $N_\mathrm{S}$ NDA ports. Let the vector $\mathbf{c}\in\mathbb{C}^{N_\mathrm{S}\times 1}$ contain the $N_\mathrm{S}$ load impedances, and let the vector $\mathbf{r}\in\mathbb{C}^{N_\mathrm{S}\times 1}$ contain the corresponding $N_\mathrm{S}$ reflection coefficients of the loads. The $i$th entry of $\mathbf{c}$, $c_i$, is related to the $i$th entry of $\mathbf{r}$, $r_i$, as follows: $r_i = (c_i-Z_0)/(c_i+Z_0)$. Well-established expressions exist for $\hat{\mathbf{S}}$ as a function of $\mathbf{S}$ and $\mathbf{r}$~\cite{anderson_cascade_1966,ha1981solid,ferrero1992new,reveyrand2018multiport,prod2024efficient}:
\begin{equation}
    \hat{\mathbf{S}}(\mathbf{r}) = \mathbf{S}_\mathcal{AA} + \mathbf{S}_\mathcal{AS} \left( \left[ \mathrm{diag}(\mathbf{r}) \right]^{-1} -  \mathbf{S}_\mathcal{SS} \right)^{-1}   \mathbf{S}_\mathcal{SA},
    \label{eq_load_S}
\end{equation}
as well as for $\hat{\mathbf{Z}}= Z_0 (\mathbf{I}_{N_\mathrm{A}} + \hat{\mathbf{S}}) (\mathbf{I}_{N_\mathrm{A}} - \hat{\mathbf{S}})^{-1}$ as a function of $\mathbf{Z}$ and $\mathbf{c}$~\cite{reveyrand2018multiport,tapie2023systematic,del2023ris,prod2024efficient}:
\begin{equation}
        \hat{\mathbf{Z}}(\mathbf{c}) = {\mathbf{Z}}_\mathcal{AA} - {\mathbf{Z}}_\mathcal{AS} \left( {\mathbf{Z}}_\mathcal{SS} + \mathrm{diag}(\mathbf{c})\right)^{-1} {\mathbf{Z}}_\mathcal{SA}.
    \label{Zaug}
\end{equation}

Our goal can now be restated as follows: By measuring $\hat{\mathbf{S}}(\mathbf{r})$ (or $\hat{\mathbf{Z}}(\mathbf{c})$) for different realizations of $\mathbf{r}$ (or $\mathbf{c}$), we seek to estimate $\mathbf{S}$ (or $\mathbf{Z}$) \textit{with minimal ambiguity}.

\subsection{Connection to Other Topics of Contemporary Interest}
\label{Sec_detailedContextualization}

We already elaborated on the relevance of the considered problem for the metrology of circuits and antenna systems in the introduction. Here, we discuss the connection to two contemporary topics that are not traditionally associated with metrology. However, as we explain, they turn out to be limited versions of the problem we tackle in our paper because they only seek to estimate parts of $\mathbf{S}$ and they tolerate ambiguities. These easier problems are already solved, as we highlight, but it is worthwhile to understand their connection to the present work, not least to clarify why these existing solutions do not solve the problem at hand in this paper.

\subsubsection{Optimal non-invasive blind focusing on a perturbation-inducing target inside an unknown complex medium}
\label{subsubsec_foc}
  
Is it possible to retrieve a vector collinear with the transmission vector from a set of sources to a target embedded inside an unknown complex medium purely based on how the target perturbs the system's measurable scattering or transmission matrix? If yes, then optimal focusing on the target is possible by phase conjugation without knowing where the target is located (``blind'') and without invasively measuring the sought-after transmission vector by attaching a transmission line to the target (``non-invasively''). Modulations of the target port's termination between different loads or displacements of the target can be the origin of such target-induced perturbations, serving as natural guidestar. Incidentally, the same problem could also be recast as pilot-exchange-free conjugate beamforming to a user equipment in a context of wireless communications. 

The problem can be solved most clearly and effectively based on a singular value decomposition (SVD) of the change of the measurable scattering matrix due to the target perturbation, requiring at least one switch of load impedance or at least two displacements~\cite{sol2024optimal}.\footnote{Earlier attempts to solve the problem can be found in Refs.~\cite{abboud2013noniterative,ma2014time,zhou2014focusing,ruan2017focusing,ambichl2017focusing,horodynski2020optimal,del2021coherent,yeo2022time}. The wavefront that optimally focuses on the target is also optimal to distinguish between two terminations of the target's port~\cite{bouchet2021optimal}.}
The problem is limited to $N_\mathrm{S}=1$ and retrieving a vector collinear with (but not necessarily equal to) $\mathbf{S}_\mathcal{AS}$. Hence, $\mathbf{S}_\mathcal{AA}$ and $\mathbf{S}_\mathcal{SS}$ are not retrieved, and the ambiguity of $\mathbf{S}_\mathcal{AS}$ is not lifted.

The question tackled in the present paper aims to establish a \textit{multitude} of load-tunable target ports inside a complex medium as ``virtual ports'' to the system. Specifically, this work seeks to retrieve \textit{with minimal ambiguity} the reflection coefficients of these ``virtual ports'' as well as the transmission coefficients between pairs of ``virtual ports'' as well as between pairs of ``virtual'' and actual ports, promising non-invasive protocols for sensing, imaging, communications and wireless power transfer in complex media. Thereby, we go significantly beyond efforts only aimed at retrieving a vector collinear with the transmission coefficients from the actual ports to a single target, as done for blind optimal focusing on a perturbation-inducing target in Ref.~\cite{sol2024optimal} and related prior works~\cite{abboud2013noniterative,ma2014time,zhou2014focusing,ruan2017focusing,ambichl2017focusing,horodynski2020optimal,del2021coherent,yeo2022time}.

\subsubsection{Physics-compliant end-to-end channel estimation in RIS-parametrized unknown rich-scattering ``smart radio environments''}

Is it possible to physics-compliantly estimate how the configuration of a reconfigurable intelligent surface (RIS) parametrizes an end-to-end wireless channel in a rich-scattering ``smart radio environment''~\cite{Liaskos_Visionary_2018,di2020smart,alexandropoulos2021reconfigurable}?\footnote{Conceptually related massively tunable complex systems are also emerging in optics~\cite{resisi2020wavefront,eliezer2023tunable,li2023adaptive}, nanophotonics~\cite{bruck2016all,shen2017deep,bogaerts2020programmable,delaney2021nonvolatile,wu2024freeform} and room acoustics~\cite{ma2018shaping,wang2022controlling}.} The RIS elements are antenna elements whose ports are terminated by tunable load impedances~\cite{tapie2023systematic,del2023ris}.\footnote{Earlier \textit{theoretical} works already used multi-port network or coupled-dipole representations but neglected the structural scattering of antennas and RIS elements and assumed that the radio environment was simply free space (see Ref.~\cite{nerini2023universal} and references therein) or considered contrived radio environments composed of discrete dipoles with known locations and properties~\cite{faqiri2022physfad,prod2023efficient,rabault2023tacit,mursia2023saris}. The important insight that no explicit description of the radio environment or structural scattering is necessary because the entries of $\mathbf{Z}$ or $\mathbf{S}$ lump together all coupling effects between the $N_\mathrm{A}+N_\mathrm{S}$ ports of interest, implying that the number of model parameters does not depend on the radio environment's complexity, was first pointed out and leveraged in Refs.~\cite{sol2023experimentally,tapie2023systematic,del2023ris}.} 
In a numerical study, a single simulation can yield $\mathbf{Z}$ or $\mathbf{S}$ free of any ambiguity, irrespective of the environment's complexity~\cite{tapie2023systematic}. Experimentally, however, the RIS ports cannot be accessed to inject or receive waves, and a detailed description of the unknown rich-scattering radio environment is unavailable.

The problem was solved by estimating via gradient descent the parameters of a physics-compliant coupled-dipole model (in this context functionally comparable to the scattering parameter representation) based on measurements of the measurable scattering matrix for known RIS configurations~\cite{sol2023experimentally}. The configurations of the 1-bit programmable RIS were known in the form of a Boolean vector $\mathbf{b}$; the values of the corresponding load impedance vector $\mathbf{c}$ were not known.
The problem resembles the one we tackle in the present paper except for the important difference that it does not care to remove ambiguities in the estimated parameters since it only matters to correctly predict the measurable scattering matrix. In the present paper, we seek to sufficiently constraint the parameter estimation problem to minimize ambiguities. Moreover, besides a gradient-descent approach, we seek a closed-form approach to gain insights into basic requirements to minimize ambiguities.

\section{Theory and Methods}
\label{Sec_theory_methods}

In this section, we develop the theory and methods for our three approaches to tackle the posed problem: one closed-form approach (based on linear algebra), and two gradient-descent approaches (one for complex-valued measurements and one for intensity-only measurements). The three approaches are complementary: the closed-form approach provides insights into fundamental requirements to minimize ambiguities, the gradient-descent approach provides flexibility and robustness against measurement noise, and the intensity-only gradient-descent approach alleviates the detection hardware cost. We assume reciprocal DUTs in this paper and leave an extension to non-reciprocal DUTs for future work.

\subsection{Closed-Form Approach}
\label{subsec_CLFO}

Our closed-form approach is derived from Eq.~(\ref{Zaug}) in terms of impedance parameters in this section but could equivalently be developed in terms of scattering parameters based on  Eq.~(\ref{eq_load_S}). We start by considering a special case in which one of the available loads at each NDA port is an ideal open-circuit (OC) termination in Sec.~\ref{subsubsec_special}; this simplifies the closed-form analysis in terms of impedance parameters. We subsequently provide a generalization to scenarios in which none of the available loads emulates any specific calibration standard in Sec.~\ref{subsubsec_general}. 

Our closed-form approach is by construction matrix-valued and identical for any value of $N_\mathrm{A}$ greater than unity. Thereby, no additional complexity arises from trying to make use of more than one or two available accessible ports. This is important because more accessible ports help to mitigate the influence of measurement noise.

The presented closed-form approach yields three fundamental insights:
\begin{enumerate}
    \item At least three \textit{distinct} and \textit{known} individual loads must be available at each NDA port.
    \item Inevitable sign ambiguities on off-diagonal scattering (or impedance) coefficients associated with NDA ports exist.
    \item Under ideal (low-noise) conditions, the upper bound on the number of required load configurations is $1+2N_\mathrm{S}+N_S(N_S-1)/2$ if $N_\mathrm{A}>1$ and  $1+2N_\mathrm{S}+N_S(N_S-1)$ if $N_\mathrm{A}=1$. 
\end{enumerate}
The basic requirement in terms of available load states is hence inherently not satisfied by a 1-bit programmable RIS (even if the two available RIS element states are characterized). These first two fundamental insights echo findings in earlier related works~\cite{garbacz1964determination,bauer1974embedding,mayhan1994technique,davidovitz1995reconstruction,pfeiffer2005recursive,pfeiffer2005equivalent,pfeiffer2005characterization,pursula2008backscattering,bories2010small,van2020verification,sahin2021noncontact,kruglov2023contactless,wiesbeck1998wide,monsalve2013multiport} which differ from the present paper in that they considered limited or different versions of the problem stated in Sec.~\ref{subsec_ProblemStatement}, often in special cases with a priori knowledge. Moreover, to the best of our knowledge, none of these earlier works includes a similar matrix-valued approach applicable to any $N_\mathrm{A}>1$.

\subsubsection{Special Case: OC Load is Available}
\label{subsubsec_special}

By inspection of Eq.~(\ref{Zaug}), the OC load plays a special role when working with impedance parameters: If all NDA ports are terminated with OC ($c_\mathrm{OC}=\infty$), the measurable $\hat{\mathbf{Z}}$ equals $\mathbf{Z}_\mathcal{AA}$.\footnote{The matched load plays a similar role when working with scattering parameters: If all NDA ports are terminated with a matched load, the measurable $\hat{\mathbf{S}}$ equals $\mathbf{S}_\mathcal{AA}$.} For this reason, we conveniently use the OC load as default load for the NDA ports in this subsection.

First, we consider the case in which all NDA ports are terminated with OC, allowing us to directly measure $\mathbf{Z}_\mathcal{AA}$. 

Second, we consider the case in which all NDA ports except for the $i$th one are terminated with OC, in which case the impedance matrix has the following structure:
\begin{equation}
    \mathbf{Z}_i = \begin{bmatrix}  \mathbf{Z}_\mathcal{AA}  &  \mathbf{z}_i \\ \mathbf{z}_i^T & \zeta_i    \\ \end{bmatrix},
    \label{eq_S1}
\end{equation}
where $\mathbf{z}_i = \mathbf{Z}_{\mathcal{A}\mathcal{S}_i} \in\mathbb{C}^{N_\mathrm{A}\times 1}$ and $\zeta_i= \mathbf{Z}_{\mathcal{S}_i\mathcal{S}_i}\in\mathbb{C}^{1\times 1}$. Switching the load impedance $c_i$ of the $i$th NDA port between two distinct values will result in a rank-one update of $\hat{\mathbf{Z}}_i= {\mathbf{Z}}_\mathcal{AA} - \mathbf{z}_i \left( \zeta_i + c_i\right)^{-1} \mathbf{z}_i^T $ such that the first and only significant singular vector $\tilde{\mathbf{z}}_i$ of the change of $\hat{\mathbf{Z}}_i$ must be collinear with $\mathbf{z}_i$: $\mathbf{z}_i=\beta_i\tilde{\mathbf{z}}_i$, where $\beta_i$ is a complex-valued scalar. This is analogous to the problem of focusing on a perturbation-inducing target~\cite{sol2024optimal} except that we work with an impedance matrix rather than scattering matrix here. The difference between the measurable impedance matrices $\hat{\mathbf{Z}}_i^\mathrm{A}$ and $\hat{\mathbf{Z}}_i^\mathrm{B}$, corresponding to terminating the $i$th NDA port with load impedance $c_i^\mathrm{A}$ or $c_i^\mathrm{B}$ and the remaining NDA ports with OC, must be of rank one and satisfy the following relation:
\begin{equation}
    \Delta\hat{\mathbf{Z}}_i^{\mathrm{AB}} = k_i^{\mathrm{AB}} \tilde{\mathbf{z}}_i \tilde{\mathbf{z}}_i^T = \beta_i^2\left[ \left(\zeta_i + c_i^\mathrm{B} \right)^{-1} - \left(\zeta_i + c_i^\mathrm{A} \right)^{-1}  \right] \tilde{\mathbf{z}}_i \tilde{\mathbf{z}}_i^T,
    \label{eq5}
\end{equation}
where $k_i^{\mathrm{AB}}$ is a complex-valued scalar that can be straight-forwardly determined given $ \Delta\hat{\mathbf{Z}}_i^{\mathrm{AB}}$ and $\tilde{\mathbf{z}}_i$.
A single switch of the $i$th load impedance value is hence insufficient to determine the two unknowns $\zeta_i$ and $\beta_i$ without ambiguity. However, as detailed in Appendix~\ref{Appendix_eq6}, upon switching between three distinct load impedances $c_i^\mathrm{A}$, $c_i^\mathrm{B}$ and $c_i^\mathrm{C}$, one can determine the values of $\zeta_i$ and $\beta_i^2$ without ambiguity:
\begin{subequations}
    \begin{equation}
         \zeta_i = \frac{ k_i^{\mathrm{AB}}c_i^\mathrm{B}\left(c_i^\mathrm{C}-c_i^\mathrm{A}\right) - k_i^{\mathrm{AC}}c_i^\mathrm{C}\left(c_i^\mathrm{B}-c_i^\mathrm{A}\right) }{k_i^{\mathrm{AC}}\left(c_i^\mathrm{B}-c_i^\mathrm{A}\right) - k_i^{\mathrm{AB}}\left(c_i^\mathrm{C}-c_i^\mathrm{A}\right)}.
    \label{eq__6a}
    \end{equation}
    \begin{equation}
         \beta_i^2 = - k_i^{\mathrm{AB}} \left[\left(\zeta_i + c_i^\mathrm{A} \right)^{-1} - \left(\zeta_i + c_i^\mathrm{B} \right)^{-1} \right]^{-1}.
    \label{eq__6b}
    \end{equation}
    \label{eq__6}
\end{subequations}

Given that Eq.~(\ref{eq5}) features only $\beta_i^2$ which is fundamentally insensitive to the sign of $\beta_i$, an ambiguity about the sign of $\beta_i$ manifests itself that cannot be resolved based on the considered type of measurements. Even though we focus on reciprocal systems in the present paper, we can easily see that the sign ambiguity also exists in non-reciprocal systems. In the latter, instead of $\beta_i^2=(-\beta_i)^2$ we would have $\beta_{i,\mathcal{AS}}\beta_{i,\mathcal{SA}} = (-\beta_{i,\mathcal{AS}})(-\beta_{i,\mathcal{SA}})$. 
In Sec.~\ref{sec_discuss_ambiguity}, we discuss in what applications scenarios this sign ambiguity is (un)problematic as well as various possibilities to lift this sign ambiguity based on modifications of the problem statement from Sec.~\ref{subsec_ProblemStatement}, either involving additional a priori knowledge about the DUT or additional allowed types of terminations of NDA ports or modified access rules for NDA ports.

Third, after having conducted the above procedure for each NDA port, we consider the case in which all NDA ports except for the $i$th and $j$th ones are terminated with OC (for $i\neq j$), in which case the impedance matrix has the following structure:
\begin{equation}
    \mathbf{Z}_{ij} = \begin{bmatrix}  \mathbf{Z}_\mathcal{AA}  &  \mathbf{z}_i &  \mathbf{z}_j \\ \mathbf{z}_i^T & \zeta_i & \kappa_{ij}   \\ \mathbf{z}_j^T  & \kappa_{ji} & \zeta_j   \\ \end{bmatrix},
    \label{eq7}
\end{equation}
where $\kappa_{ij}=\mathbf{Z}_{\mathcal{S}_i\mathcal{S}_j}=\mathbf{Z}_{\mathcal{S}_j\mathcal{S}_i}=\kappa_{ji}\in\mathbb{C}^{1\times 1}$ is the only remaining unknown. Switching from $c_i=c_\mathrm{OC}$ and $c_j=c_\mathrm{OC}$ to $c_i^\mathrm{B}\neq c_\mathrm{OC}$ and $c_j^\mathrm{B}\neq c_\mathrm{OC}$ results in a rank-two update $\mathbf{D}\in\mathbb{C}^{N_\mathrm{A} \times N_\mathrm{A}  }$ of the measurable impedance matrix $\hat{\mathbf{Z}}_{ij}$. As detailed in Appendix~\ref{Appendix_eq9}, 
\begin{equation}
 {\mathbf{Z}}_\mathcal{A\bar{S}}^+  \mathbf{D} {\mathbf{Z}}_\mathcal{\bar{S}A}^+ = \frac{1}{(\zeta_i+c_i^\mathrm{B})(\zeta_j+c_j^\mathrm{B}) - \kappa_{ij}^2}        \begin{bmatrix}    \zeta_j+c_j^\mathrm{B} & -\kappa_{ij} \\ -\kappa_{ij} &  \zeta_i+c_i^\mathrm{B}     \\ \end{bmatrix},
    \label{eq8}
\end{equation}
where ${\mathbf{Z}}_\mathcal{A\bar{S}}^+$ denotes the Moore–Penrose pseudo-inverse of ${\mathbf{Z}}_\mathcal{A\bar{S}} = \begin{bmatrix}    \mathbf{z}_i &  \mathbf{z}_j \end{bmatrix} $ and $\mathcal{\bar{S}} = \{i,j\}$. Assuming $N_\mathrm{A}>1$, straightforward algebraic manipulations detailed in Appendix~\ref{Appendix_eq9} yield
\begin{equation}
    \kappa_{ij} = -\left[{\mathbf{Z}}_\mathcal{A\bar{S}}^+  \mathbf{D} {\mathbf{Z}}_\mathcal{\bar{S}A}^+\right]_{12} \left( (\zeta_i+c_i^\mathrm{B})(\zeta_j+c_j^\mathrm{B}) - \kappa_{ij}^2  \right),
    \label{eq9}
\end{equation}
where
\begin{equation}
     \kappa_{ij}^2 = (\zeta_i+c_i^\mathrm{B})(\zeta_j+c_j^\mathrm{B}) - \frac{\zeta_j + c_j^\mathrm{B}}{\left[{\mathbf{Z}}_\mathcal{A\bar{S}}^+  \mathbf{D} {\mathbf{Z}}_\mathcal{\bar{S}A}^+\right]_{11}}.
     \label{eq10}
\end{equation}

The matrix-valued quadratic equation for $\kappa_{ij}$ defined in Eq.~(\ref{eq_S16}) can be solved for $\kappa_{ij}$ only if $N_\mathrm{A} > 1$. If $N_\mathrm{A} = 1$, two distinct realizations of terminations other than OC for the $i$th and $j$th port are required to retrieve $\kappa_{ij}$. Hence, the use of at least two accessible ports reduces the number of required configurations by $N_S(N_S-1)/2$. Moreover, the larger $N_\mathrm{A}$ is, the better the robustness against measurement noise will be.

\begin{algorithm}[h]
	\caption{Closed-form algorithm}
	\label{Alg1}
	\KwIn{Measurements of $\hat{\mathbf{Z}}$ for \textit{specific} load impedance configurations $\mathbf{c}$ (see left frame in Fig.~\ref{Fig0}B).}
    Estimate $\mathbf{Z}_\mathcal{AA}$.\\
	\For{$i=1,2,\ldots,N_{\rm S}$}{
        Estimate $\tilde{\mathbf{z}}_i$.\\
        Estimate $\zeta_i$ and $\beta_i^2$.\\
    }
	\For{$i=1,2,\ldots,N_{\rm S}$}{
        \For{$j=1,2,\ldots,N_{\rm S}$}{
            \If{$i>j$}     {Estimate $\kappa_{ij}$.}    
        }
    } 
    \textbf{(Optional)} Convert the estimate of $\mathbf{Z}$ to $\mathbf{S}$. \\
	\KwOut{Estimate of $\mathbf{Z}$ (and/or $\mathbf{S}$).}
\end{algorithm}

There exists hence a closed-form approach, summarized in Algorithm~\ref{Alg1}, to estimate all entries of $\mathbf{Z}$ despite $N_\mathrm{S}$ NDA ports for which only the termination with tunable individual loads is possible -- except for the sign ambiguity on the off-diagonal entries involving one or two NDA port(s). The estimated $\mathbf{Z}$ can be converted to $\mathbf{S}$ which fill feature the same sign ambiguities as $\mathbf{Z}$ for the off-diagonal entries associated with NDA ports.

\subsubsection{General Case: OC Load is NOT Available}
\label{subsubsec_general}

In this subsection, we provide a procedure to generalize the approach from Sec.~\ref{subsubsec_special} to cases in which no ideal OC loads are available. In such cases, instead of $c_\mathrm{OC}$, we use some $c_{0,i}$ as default load for the $i$th NDA port. 

\begin{figure}[h]
    \centering
    \includegraphics[width=\columnwidth]{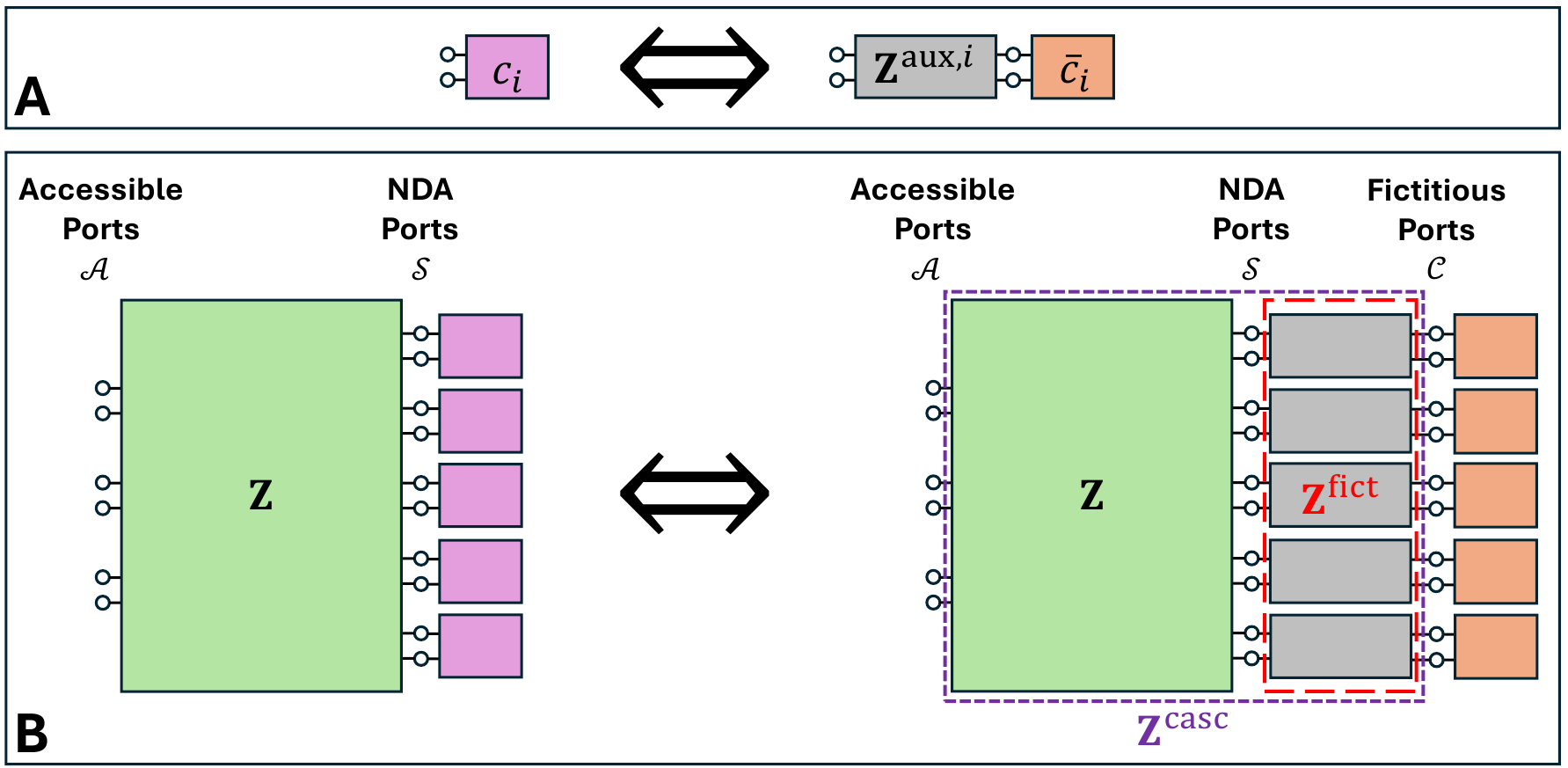}
    \caption{(A) Illustration of an equivalent representation of a load $c_i$ (purple) in terms of a two-port auxiliary system (grey, characterized by its impedance matrix $\mathbf{Z}^{\mathrm{aux},i}$) terminated by a different load impedance $\bar{c}_i$ (orange). Each port is represented by a pair of two terminals. (B) Application of the concept from (A) to all loads terminating the NDA ports of the main system. A fictitious system (characterized by its impedance matrix $\mathbf{Z}^\mathrm{fict}$) comprising the $N_\mathrm{S}$ auxiliary two-port systems is defined by the red dashed box. The purple dashed box defines the cascade of the main system and the fictitious system. The impedance matrix characterizing this cascade is $\mathbf{Z}^\mathrm{casc}$.}
    \label{Fig1bis}
\end{figure}

First, as illustrated in Fig.~\ref{Fig1bis}A, we recognize that we can represent any one-port system with impedance $c_i$ as an auxiliary two-port system terminated by a different one-port system with impedance $\bar{c}_i$. With
\begin{equation}
    \mathbf{Z}^{\mathrm{aux},i} = \begin{bmatrix} Z^{\mathrm{aux},i}_{11} & Z^{\mathrm{aux},i}_{12}\\ Z^{\mathrm{aux},i}_{21} & Z^{\mathrm{aux},i}_{22}  \end{bmatrix} \in \mathbb{C}^{2 \times 2},
\end{equation}
being the impedance matrix of the auxiliary system, the application of Eq.~(\ref{Zaug}) yields:
\begin{equation}
    c_i = Z^{\mathrm{aux},i}_{11} - \frac{Z^{\mathrm{aux},i}_{12}  Z^{\mathrm{aux},i}_{21}}{Z^{\mathrm{aux},i}_{22} + \bar{c}_i}.
    \label{eq__12}
\end{equation}
Applied to our actual and desired default loads at the $i$th NDA port, this yields:
\begin{equation}
    c_{0,i} = Z^{\mathrm{aux},i}_{11} - \frac{Z^{\mathrm{aux},i}_{12}  Z^{\mathrm{aux},i}_{21}}{Z^{\mathrm{aux},i}_{22} + c_\mathrm{OC}} = Z^{\mathrm{aux},i}_{11},
\end{equation}
revealing that we must choose $Z^{\mathrm{aux},i}_{11} =  c_{0,i}$ and can freely select arbitrary non-zero values for $Z^{\mathrm{aux},i}_{12}=Z^{\mathrm{aux},i}_{21}$ and $Z^{\mathrm{aux},i}_{22}$, where we impose for simplicity that the auxiliary two-port system is reciprocal.

Second, for the other two non-default loads available at the $i$th port, we can identify an equivalent representation based on the same auxiliary two-port system by rearranging Eq.~(\ref{eq__12}):
\begin{equation}
    \bar{c}_i = \frac{Z^{\mathrm{aux},i}_{12}Z^{\mathrm{aux},i}_{21}}{Z^{\mathrm{aux},i}_{11} - c_i} - Z^{\mathrm{aux},i}_{22}.
\end{equation}

Third, the above steps are repeated for each NDA port. There is hence no requirement that the non-OC default loads of the NDA ports are identical.

Fourth, as illustrated in Fig.~\ref{Fig1bis}B, we can now represent our main problem, involving available loads $c_i^\mathrm{A}$, $c_i^\mathrm{B}$ and $c_i^\mathrm{C}$ at the $i$th NDA port (of which none equals $c_\mathrm{OC}$ and one is arbitrarily chosen to be $c_{0,i}$), in terms of the determined auxiliary two-port systems ($\mathbf{Z}^{\mathrm{aux},i}$ for the $i$th NDA port) and the three loads $\bar{c}_i^\mathrm{A}$, $\bar{c}_i^\mathrm{B}$ and $\bar{c}_i^\mathrm{C}$ at the $i$th NDA port (of which one equals $c_\mathrm{OC}$). As highlighted by the red dashed box in Fig.~\ref{Fig1bis}B, it is convenient to summarize all two-port auxiliary systems in a fictitious supersystem:
\begin{equation}
    \mathbf{Z}^\mathrm{fict} = \mathrm{blockdiag}(\mathbf{Z}^{\mathrm{aux},i}) \in \mathbb{C}^{2N_\mathrm{S} \times 2 N_\mathrm{S}}.
\end{equation}

Now, we can apply the method from Sec.~\ref{subsubsec_special} using the measured impedance matrices but assuming the applied load impedances were $\bar{c}_i^\mathrm{A}$, $\bar{c}_i^\mathrm{B}$ and $\bar{c}_i^\mathrm{C}$ instead of  $c_i^\mathrm{A}$, $c_i^\mathrm{B}$ and $c_i^\mathrm{C}$ at the $i$th NDA port. Thereby, we will reconstruct an impedance matrix $\mathbf{Z}^\mathrm{casc}\in\mathbb{C}^{N \times N}$ that is \textit{not} the sought-after impedance-matrix $\mathbf{Z}$ but rather the cascade of $\mathbf{Z}$ with $\mathbf{Z}^\mathrm{fict}$, as illustrated in Fig.~\ref{Fig1bis}B by the purple dashed box. There is a sign ambiguity on the off-diagonal entries of the estimated $\mathbf{Z}^\mathrm{casc}$ associated with fictitious ports (see Fig.~\ref{Fig1bis}B), for the same reason that there was a sign ambiguity on the off-diagonal entries of $\mathbf{Z}$ associated with NDA ports in Sec.~\ref{subsubsec_special}.

The last, fifth, step consists hence in determining $\mathbf{Z}$ given $\mathbf{Z}^\mathrm{casc}$ and $\mathbf{Z}^\mathrm{fict}$; $\mathbf{Z}^\mathrm{fict}$ is known perfectly. The required mathematical procedure determines the four blocks of $\mathbf{Z}$ as follows~\cite{Jabotinski2018efficient}:\footnote{Eq.~(\ref{eq_17_}) can be understood as the equivalent of an inverse Redheffer star product with impedance instead of scattering parameters.}
\begin{equation}
\begin{split}
& \mathbf{Z}_\mathcal{AA} = \mathbf{Z}^\mathrm{casc}_\mathcal{AA} + \mathbf{Z}^\mathrm{casc}_\mathcal{AC} \mathbf{W} \mathbf{Z}^\mathrm{casc}_\mathcal{CA}\\
& \mathbf{Z}_\mathcal{AS} = \mathbf{Z}^\mathrm{casc}_\mathcal{AC} \mathbf{W} \mathbf{Z}^\mathrm{fict}_{\mathcal{CS}} \\
& \mathbf{Z}_\mathcal{SA} = \mathbf{Z}^\mathrm{fict}_{\mathcal{SC}} \mathbf{W} \mathbf{Z}^\mathrm{casc}_\mathcal{CA} \\
& \mathbf{Z}_\mathcal{SS} = -\mathbf{Z}^\mathrm{fict}_{{\mathcal{SS}}} + \mathbf{Z}^\mathrm{fict}_{\mathcal{SC}} \mathbf{W} \mathbf{Z}^\mathrm{fict}_{\mathcal{CS}}\\
\end{split}
    \label{eq_17_}
\end{equation}
and
\begin{equation}
    \mathbf{W} = \left( \mathbf{Z}^\mathrm{fict}_{{\mathcal{CC}}} - \mathbf{Z}^\mathrm{casc}_\mathcal{CC}  \right)^{-1},
    \label{eq8_}
\end{equation}
where $\mathcal{C}$ denotes the set of fictitious port indices, as shown in Fig.~\ref{Fig1bis}B.

\subsection{Gradient-Descent Approach}
\label{subsec_gradesc}

In this subsection, we develop a gradient-descent approach similar to Ref.~\cite{sol2023experimentally} (but recall that Ref.~\cite{sol2023experimentally} involved an RIS with 1-bit programmable elements whose two possible states were not characterized and Ref.~\cite{sol2023experimentally} did not seek to remove any ambiguities). This approach, summarized in Algorithm~\ref{Alg2} and detailed below, takes a set of random known load configurations $\mathbf{c}$ and corresponding measurements $\hat{\mathbf{S}}$ as input in order to retrieve all entries of $\mathbf{S}$. Of course, there is still the above identified requirement for at least three distinct and known load impedances and the fundamental sign ambiguity, but the gradient-descent approach present four practical advantages:
\begin{enumerate}
    \item The load configurations can be arbitrary (as long as they are known), implying a  compatibility with opportunistic load switches (originating, for example, from the regular operation of a backscatter communications system~\cite{liu2013ambient,zhao2020metasurface}).
    \item The changes of $\hat{\mathbf{S}}$ are larger than in the closed-form approach because subsequent load configurations typically differ regarding the termination of significantly more than one or two NDA ports (especially for large $N_\mathrm{S}$).
    \item The number of utilized measurements can be chosen flexibly. Under low-noise conditions, it can remain below the upper bound identified with the closed-form approach (this effect is more dramatic for large values of $N_\mathrm{S}$ but already appreciable in our experiment with $N_\mathrm{S}=4$ in Sec.~\ref{sec_exp}). Under noisy conditions, one can flexibly add measurements to mitigate the adverse effect of noise. 
    \item There is no additional complexity in the data analysis if none of the available loads emulates a calibration standard such as OC.
\end{enumerate}

Here, we work directly with scattering parameters, see Eq.~(\ref{eq_load_S}), and proceed in two steps. (A similar approach could also be implemented with impedance parameters but since our raw measurements are scattering parameters, this would be somewhat more cumbersome.) First, we seek to estimate the blocks $\mathbf{S}_\mathcal{AS}=\mathbf{S}_\mathcal{SA}^T$ and $\mathbf{S}_\mathcal{SS}=\mathbf{S}_\mathcal{SS}^T$. The only constraint we impose is reciprocity. Based on Eq.~(\ref{eq_load_S}), the change of $\hat{\mathbf{S}}$ due to a change of load configuration from $\mathbf{c}^\mathrm{A}$ to $\mathbf{c}^\mathrm{B}$ (corresponding to reflection coefficients $\mathbf{r}^\mathrm{A}$ and $\mathbf{r}^\mathrm{B}$, respectively) is
\begin{equation}
\begin{split}
    \Delta\hat{\mathbf{S}}^\mathrm{AB} = \mathbf{S}_\mathcal{AS} \bigg[ 
    &\left( \left[\mathrm{diag}(\mathbf{r}^\mathrm{B})\right]^{-1} - \mathbf{S}_\mathcal{SS} \right)^{-1} \\
    - &\left( \left[\mathrm{diag}(\mathbf{r}^\mathrm{A})\right]^{-1} - \mathbf{S}_\mathcal{SS} \right)^{-1} \bigg] 
    \mathbf{S}_\mathcal{SA}.
\end{split}
\label{eq11}
\end{equation}
Given $N_\mathrm{cal}$ measurements with random known load configurations\footnote{The state of each load in a random load configuration is independently selected at random with equal probability for each state.}, we use pairs of subsequent measurements to generate $N_\mathrm{cal}-1$ triplets $\{\mathbf{r}^\mathrm{A},\mathbf{r}^\mathrm{B},\Delta\hat{\mathbf{S}}^\mathrm{AB}\}$ that we use to estimate the values of  $\mathbf{S}_\mathcal{AS}$ and $\mathbf{S}_\mathcal{SS}$. The cost function to be minimized is simply the mean absolute error of the entries of $\Delta\hat{\mathbf{S}}^\mathrm{AB}$. Details about the implementation of the gradient descent are detailed in Appendix~\ref{appendixB}.

Then, we estimate $\mathbf{S}_\mathcal{AA}$ as the average of our $N_\mathrm{cal}$ realizations of
\begin{equation}
    \mathbf{S}_\mathcal{AA} = \hat{\mathbf{S}}^\mathrm{A} - \mathbf{S}_\mathcal{AS} \left( \left[\mathrm{diag}(\mathbf{r}^\mathrm{A}) \right]^{-1} -  \mathbf{S}_\mathcal{SS} \right)^{-1}   \mathbf{S}_\mathcal{SA},
    \label{eq12}
\end{equation}
where $\hat{\mathbf{S}}^\mathrm{A} $ denotes the measurable scattering matrix $\hat{\mathbf{S}}$ for the load reflection coefficients configuration $\mathbf{r}^\mathrm{A}$.

\begin{algorithm}[h]
	\caption{Gradient-descent algorithm}
	\label{Alg2}
	\KwIn{Measurements of $\hat{\mathbf{S}}^\mathrm{A}$ for \textit{arbitrary} known load reflection coefficient configurations $\mathbf{r}^\mathrm{A}$.}
    Estimate $\mathbf{S}_\mathcal{AS}$ and  $\mathbf{S}_\mathcal{SS}$ via gradient descent given $N_\mathrm{cal}-1$ triplets $\{\mathbf{r}^\mathrm{A},\mathbf{r}^\mathrm{B},\Delta\hat{\mathbf{S}}^\mathrm{AB}\}$.\\
    Estimate  $\mathbf{S}_\mathcal{AA}$.\\
    \textbf{(Optional)} Convert the estimate of $\mathbf{S}$ to $\mathbf{Z}$. \\
	\KwOut{Estimate of $\mathbf{S}$ (and/or $\mathbf{Z}$).}
\end{algorithm}

The described procedure, summarized in Algorithm~\ref{Alg2}, differs from Ref.~\cite{sol2023experimentally} regarding its two-step nature and its requirement to know the applied load characteristics. The former reduces the complexity of the gradient-descent optimization, and here we evaluate $\mathbf{S_\mathcal{AA}}$ in closed form using Eq.~(\ref{eq12}). The latter is essential to minimize ambiguities. The estimation of $\mathbf{S}_\mathcal{AA}$ is not subject to any sign ambiguity.

\subsection{Intensity-Only Gradient-Descent Approach}
\label{subsec_phaseless}

In this subsection, we adapt the gradient-descent approach from the previous subsection to work with intensity-only data. Lifting the requirement to measure phase drastically alleviates the detection hardware cost, for instance, because there would be no need for synchronization. Recently, Ref.~\cite{sol2023experimentally} demonstrated that end-to-end RIS parametrized channel estimation is possible without phase information thanks to the mathematical structure of the governing equations, but does this conclusion generalize to our present goal of minimizing ambiguities? 

The retrieval of phase information from intensity data is an active area of research in signal processing~\cite{shechtman2015phase,dong2023phase}. The usual problem is to retrieve $\mathbf{x}$ [resp.~$\mathbf{S}$] given intensity measurements $\left| \mathbf{S}\mathbf{x} \right|^2$ and knowing $\mathbf{S}$ [resp.~$\mathbf{x}$]; this usual problem statement hence assumes the ability to input/output waves via all ports. If all ports are accessible, estimating the phases of $\mathbf{S}$ given measurements of $\left| \mathbf{S}\mathbf{x} \right|^2$ corresponding to a sufficient number of known and diverse input wavefronts $\mathbf{x}$ is surprisingly simple using gradient descent (see Appendix~\ref{appendixC}).\footnote{Conceptually related works in optics consider a transmission matrix $\mathbf{T}$ (i.e., an off-diagonal block of $\mathbf{S}$) instead of $\mathbf{S}$~\cite{dremeau2015reference,caramazza2019transmission}, resulting in ambiguities about the relative phases between different rows of $\mathbf{T}$ that require tunable interferences of the outgoing wavefronts to be resolved~\cite{goel2023referenceless}. In contrast, since we input and output waves via the same ports when estimating $\mathbf{S}$, there are no row-wise phase ambiguities.} Only an arbitrary global phase factor $\theta$ cannot be determined without ambiguity because $\left| \mathbf{S}\mathbf{x} \right|^2 = \left| \mathrm{e}^{\jmath\theta}\mathbf{S}\mathbf{x} \right|^2 $. Random or pseudo-random input wavefronts $\mathbf{x}$ work well, the only input wavefronts to avoid are those with one-hot encoding for which all entries except one are zero such that interferences between signals injected via different input ports cannot be probed.

Our problem, however, is much more complicated than the usual phase-retrieval problem because of the NDA load-tunable ports. Nonetheless, applying the above phase-retrieval insights, we can at least straightforwardly retrieve the phases of each $\hat{\mathbf{S}}^\mathrm{A}$ up to some global phase factor $\theta^\mathrm{A}$ that will be unknown and different for each $\hat{\mathbf{S}}^\mathrm{A}$. However, to meaningfully compute $\Delta\hat{\mathbf{S}}^\mathrm{AB}$ (only subject to a global phase ambiguity) would require $\theta^\mathrm{A}=\theta^\mathrm{B}$. Exploiting the fact that we know the rank of $\Delta\hat{\mathbf{S}}^\mathrm{AB}$ (it is equal to the number of entries by which $\mathbf{r}^\mathrm{A}$ and $\mathbf{r}^\mathrm{B}$ differ), we can easily adjust $\theta^\mathrm{B}$ until the rank of $\Delta\hat{\mathbf{S}}^\mathrm{AB}$ is the expected one (for $\theta^\mathrm{A}\neq\theta^\mathrm{B}$, the rank of $\Delta\hat{\mathbf{S}}^\mathrm{AB}$ would exceed the expected rank). The only requirement for this procedure to work is that the rank of $\Delta\hat{\mathbf{S}}^\mathrm{AB}$ is less than $N_\mathrm{S}$. Technical details are summarized in Appendix~\ref{appendixC}. By applying this approach to different quadruples $\{ \hat{\mathbf{S}}^\mathrm{A}, \mathbf{r}^\mathrm{A}, \hat{\mathbf{S}}^\mathrm{B}, \mathbf{r}^\mathrm{B} \}$ we can ensure that the unknown global phase offset is the same for all considered load configurations: $\theta = \theta^\mathrm{A}=\theta^\mathrm{B}=\dots$.

Next, we can implement the same gradient descent as in the previous section to obtain estimates of $\mathbf{S}_\mathcal{AS}$ and $\mathbf{S}_\mathcal{SS}$. In addition to the sign ambiguity on off-diagonal entries related to NDA ports, this time there is an additional blockwise phase ambiguity on $\mathbf{S}_\mathcal{AS}$ (but not on $\mathbf{S}_\mathcal{SS}$). Indeed, by inspection of Eq.~(\ref{eq11}), instead of $\mathbf{S}_\mathcal{AS}$, we expect to obtain $\mathrm{e}^{\j\theta/2}\mathbf{S}_\mathcal{AS}$ (without knowing the value of $\theta$). 
Finally, we can estimate $\mathbf{S}_\mathcal{AA}$ using Eq.~(\ref{eq12}) as in the previous section. This time, we expect a blockwise phase offset of $\theta$ on $\mathbf{S}_\mathcal{AA}$ because of the blockwise phase offset of $\theta/2$ on $\mathbf{S}_\mathcal{AS}=\mathbf{S}_\mathcal{SA}^T$. 
Given the $\theta$ ambiguity, we may as well alleviate the burden of the gradient descent optimization by defining a cost function that is insensitive to the value of $\theta$, as detailed in Appendix~\ref{appendixC}.
Due to the additional blockwise phase ambiguity in the case of intensity-only measurements, a conversion of the ambiguous estimate of $\mathbf{S}$ to $\mathbf{Z}$ is not meaningful (i.e., it does \textit{not} yield an estimate of $\mathbf{Z}$ with the same blockwise phase ambiguity and off-diagonal sign ambiguities).

\begin{algorithm}[h]
	\caption{Phaseless gradient-descent algorithm}
	\label{Alg3}
	\KwIn{Measurements of $\left|\hat{\mathbf{S}}^\mathrm{A} \hat{\mathbf{x}} \right|^2$ for \textit{arbitrary} known input wavefronts $\hat{\mathbf{x}}\in\mathbb{C}^{N_\mathrm{A}\times 1}$ for \textit{arbitrary} known load reflection coefficient configurations $\mathbf{r}^\mathrm{A}$.}
    \For{$i=1,2,\ldots,N_{\rm cal}$}{
    Estimate the phases of the $i$h realization of $\hat{\mathbf{S}}^\mathrm{A}$.\\
    }
    \For{$i=2,\ldots,N_{\rm cal}$}{
    Adjust the global phase ambiguity of the $i$th realization of $\hat{\mathbf{S}}^\mathrm{A}$. \\
    }    
    Estimate $\mathbf{S}$ via gradient descent using Algorithm~\ref{Alg2} given $N_\mathrm{cal}-1$ triplets $\{\mathbf{r}^\mathrm{A},\mathbf{r}^\mathrm{B},\Delta\hat{\mathbf{S}}^\mathrm{AB}\}$.\\
	\KwOut{Estimate of $\mathbf{S}$.}
\end{algorithm}

The presented phaseless gradient descent algorithm differs from the one used in Ref.~\cite{sol2023experimentally} in multiple important ways. In addition to those already identified in the previous section, most notably, the pre-processing steps to retrieve the phases of all $\hat{\mathbf{S}}^\mathrm{A}$ with the same global offset $\theta$ in Algorithm~\ref{Alg3} drastically reduce the burden of the main gradient descent part of Algorithm~\ref{Alg3}. These pre-processing steps are themselves very simple and can be performed in parallel for different load configurations. The relaxed cost function detailed in Appendix~\ref{appendixC} is also implemented more efficiently here.

To summarize, using the procedure outlined in this section and summarized in Algorithm~\ref{Alg3}, a limitation to phase-insensitive measurements only adds a mild blockwise phase ambiguity of $\theta/2$ [$\theta$] on $\mathbf{S}_\mathcal{AS}=\mathbf{S}_\mathcal{SA}^T$ [$\mathbf{S}_\mathcal{AA}$], revealing strong additional constraining structure in the governing equations. 
Only the gradient descent approach appears capable of coping with phase-insensitive measurements.

\section{Discussion on \\Remaining Inevitable Ambiguities}
\label{sec_discuss_ambiguity}

In the three approaches developed in the previous Sec.~\ref{Sec_theory_methods}, ambiguities remain that are inevitable given the considered problem statement from Sec.~\ref{subsec_ProblemStatement}. In all cases, there is a sign ambiguity on off-diagonal scattering coefficients associated with NDA ports. In the case of intensity-only measurements, there is an additional blockwise phase ambiguity. At this stage, it is important to discuss in what application scenarios these remaining ambiguities are (un)problematic, and what application-specific modifications of the problem statement can allow one to lift these ambiguities when they are problematic. We include this discussion on remaining inevitable ambiguities before presenting our experimental results because lifting the ambiguities facilitates evaluating and visualizing the experimental results (e.g., calculating differences between ground-truth and estimated  scattering or impedance parameters, as well as plotting the latter's phases).

\subsection{Applications in which remaining inevitable ambiguities are unproblematic}

We begin by discussing application scenarios in which the remaining inevitable ambiguities are not problematic (and hence do not require being lifted). These include scenarios in which only the diagonal entries of $\mathbf{S}_\mathcal{SS}$ or only the magnitudes of $\mathbf{S}$ are of interest.

\subsubsection{Crosstalk characterization} For the metrology of circuits and antenna systems with many ports, the key objective is oftentimes to assess whether the isolation between the DUT ports is sufficient. In that case, only the magnitudes of the estimated off-diagonal scattering coefficients matter which are not affected by any of the remaining ambiguities; hence all three presented approaches are perfectly suited to this application.

\subsubsection{Indoor surveillance and non-destructive testing} For wireless sensing applications in indoor surveillance or non-destructive testing, it is conceivable to embed wirelessly powered antennas with load-tunable ports in the walls of a room or inside a concrete structure, respectively. These would constitute the NDA ports within our framework. The retrieval of the magnitudes of the transmission coefficients between these NDA ports could enable the assessment of activities in the room or the structural integrity of the concrete, e.g., based on detecting line-of-sight blockage. Moreover, these sensor units could be powered wirelessly via phase-conjugation focusing which only requires a vector collinear to the transmission vector from the sources to the NDA ports~\cite{sol2024optimal}. Despite their remaining inevitable ambiguities, our approaches appear hence suited for such wireless sensing applications.

\subsubsection{Wireless bioelectronics}
A specific relevant example is gastrointestinal segment tracking of ingestible capsules based on how the reflection coefficients of antennas integrated into the capsules change as a function of the surrounding tissue~\cite{cil2023use}. The ports of these ingestible antennas are treated as NDA ports in our framework. Because we recover the diagonal entries of $\mathbf{S}_\mathcal{SS}$ without any ambiguities in all three methods, the presented approaches are perfectly suited to this application. Note that even for the simplest case of retrieving the reflection coefficient of a single ingestible antenna (i.e., $N_\mathrm{S}=1$), our approaches offer at least three added values compared to related existing literature~\cite{garbacz1964determination,bauer1974embedding,mayhan1994technique,davidovitz1995reconstruction,pfeiffer2005recursive,pfeiffer2005equivalent,pfeiffer2005characterization,pursula2008backscattering,bories2010small,van2020verification,sahin2021noncontact,kruglov2023contactless}: (i) we do not make any assumptions about the geometry of the setup; (ii) we can optimally incorporate information from arbitrarily many accessible ports, providing a convenient route to improved noise robustness; (iii) we can dispense with phase-sensitive detection.

\subsection{Applications in which the remaining ambiguities need to be lifted}

Of course, many examples of applications exist in which the inevitable remaining ambiguities are problematic and should be lifted. To lift them, the problem statement formulated in Sec.~\ref{subsec_ProblemStatement} needs to be modified. In the following, we discuss possible modifications of the problem statement and their suitability to various application scenarios.

\subsubsection{A priori knowledge about DUT} 

In some applications, a priori knowledge, e.g., about the DUT's characteristics near dc~\cite{lu2000port,lu2003multiport,pfeiffer2005recursive} or the DUT's geometric details, can be used to lift the ambiguities. This constitutes a modification of the problem statement from Sec.~\ref{subsec_ProblemStatement} regarding the available a priori knowledge. Whenever necessary and feasible, this is, of course, the preferable route to lifting the ambiguities because it does not involve additional measurements. However, this route is application-specific and not always possible (e.g., it is clear that we cannot have any a priori knowledge about our DUT based on a reverberation chamber shown in Fig.~\ref{Fig1_1}A).

\subsubsection{Availability of coupled loads} The use of coupled loads can lift the ambiguities, as we recently demonstrated in a follow-up paper~\cite{del2024virtual2p0} (see also related prior work in Ref.~\cite{denicke2012application}). The additional use of coupled loads constitutes a modification of the problem statement from Sec.~\ref{subsec_ProblemStatement} regarding the types of available loads (which are limited to individual one-port loads in this paper). The implementation of additional coupled loads is conceivable in many applications, as elaborated in Sec.~\ref{sec_discussion}, but not in all. For instance, connecting neighboring load-tunable antennas via coupled loads would contradict the core idea of deploying them in wireless sensing applications. At most $N_\mathrm{S}$ additional measurements are required if coupled load are used to lift the ambiguities~\cite{del2024virtual2p0}, i.e., the complexity of this technique scales linearly.

\subsubsection{Direct measurements of the transmission from one accessible to all NDA ports} 
\label{sssec_directMeas}
A single transmission measurement between one of the accessible ports and the $i$th NDA port is sufficient to lift the sign ambiguity on the $i$th column [row] in the block $\mathbf{S}_\mathcal{AS}$ [$\mathbf{S}_\mathcal{SA}$]. One simply tests with which of the two possible signs the retrieved parameters correctly predict the sign of the measured transmission coefficient.\footnote{Importantly, it is hence \textit{not} necessary to directly measure the transmission from each accessible port to each NDA port.} Once this is done for the $i$th and $j$th NDA ports, the sign ambiguity on the $(i,j)$th entry in $\mathbf{S}_\mathcal{SS}$ is also eliminated. Hence, $N_\mathrm{S}$ additional transmission measurements are required to lift the sign ambiguities with this technique, i.e., the complexity of this technique scales linearly. In the case of intensity-only measurements, the $N_\mathrm{S}$ described additional measurements (now intensity-only) eliminate both the sign and blockwise phase ambiguities.  Of course, this technique implies a clear modification of the problem statement from Sec.~\ref{subsec_ProblemStatement} regarding the definition of NDA ports because these additional measurements require sending or receiving waves once via each NDA port. Clearly, this is not viable whenever the NDA ports are embedded. However, when the NDA ports are physically accessible and only effectively NDA for practical purposes because of their large number, this technique is feasible (not least because of its linear complexity scaling). In our proof-of-principle experiment, we are in the latter situation and hence we choose this technique of ambiguity removal because of its ease of practical implementation.

\section{Experimental Results}
\label{sec_exp}

\subsection{Experimental Setup}
\label{sec_exp_setup}

\begin{figure*}
    \centering
    \includegraphics[width=\textwidth]{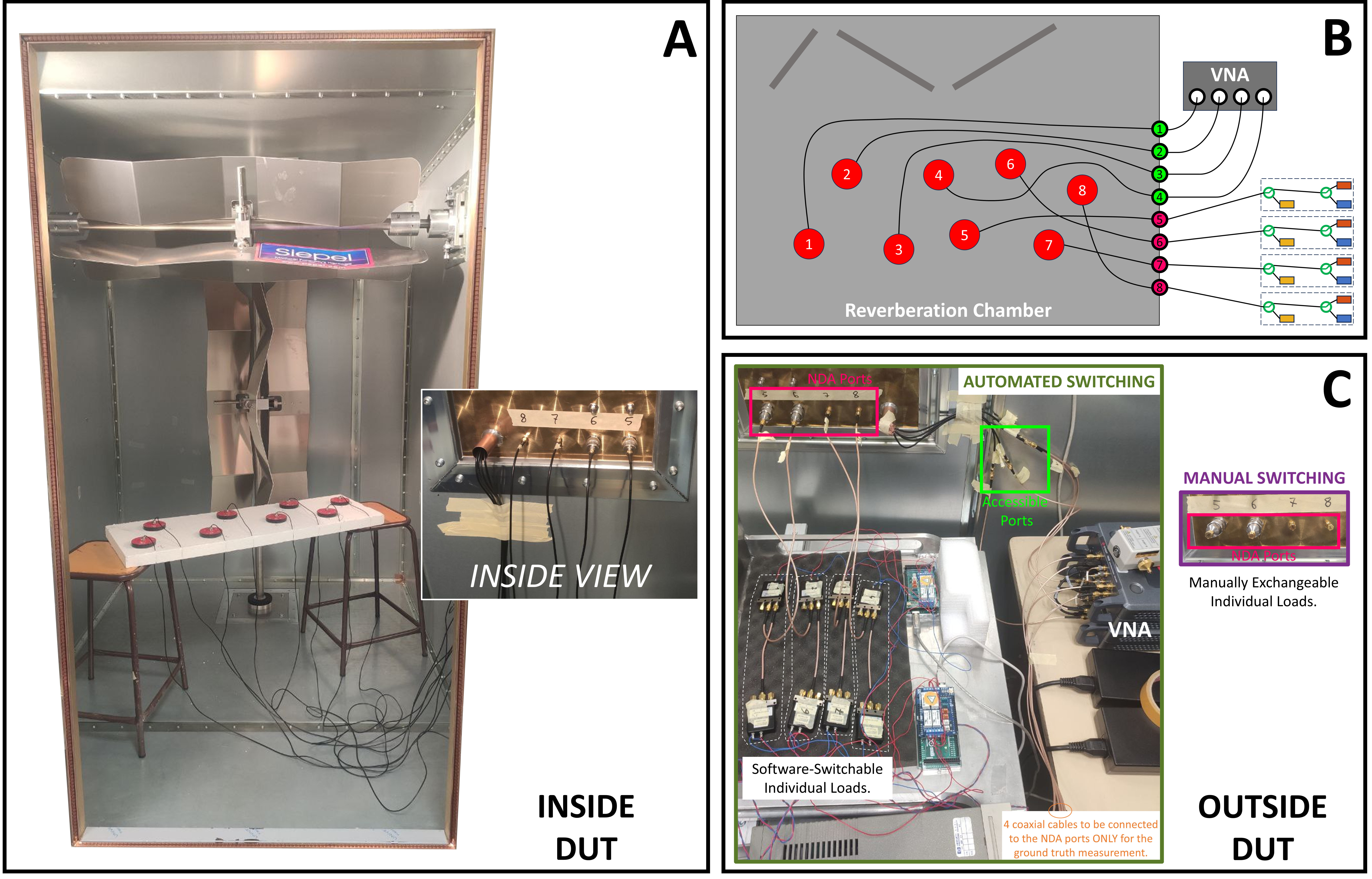}
    \caption{Experimental setup. (A) Inside view of the eight-port DUT (an RC comprising 8 antennas) with unknown geometrical and material properties. The inset shows an inside view of the four through-wall SMA mounts used as NDA ports. (B) Schematic diagram of the key elements of the experimental setup for the case of automated switching depicted in (C), excluding power and logic-control cables for the switches and the VNA. The DUT has four accessible (green) and four NDA (pink) ports. Each software-switchable load comprises two two-port electro-mechanical switches (green circles), two coaxial cables, and three distinct one-port loads (yellow, red, blue). (C) Outside view of the DUT showing that the DUT's four accessible ports (green) are directly connected to a VNA and the DUT's four NDA ports (pink) are connected to four software-switchable individual loads. On the right hand side, the alternative termination of the four NDA ports with manually exchangeable individual loads for the case of manual switching is shown. On the bottom right of the main picture, four spare coaxial cables are seen which can connect four additional VNA ports to the four NDA ports to measure the ground truth (only for validation purposes).}
    \label{Fig1_1}
\end{figure*}

\begin{figure}
    \centering
    \includegraphics[width=\columnwidth]{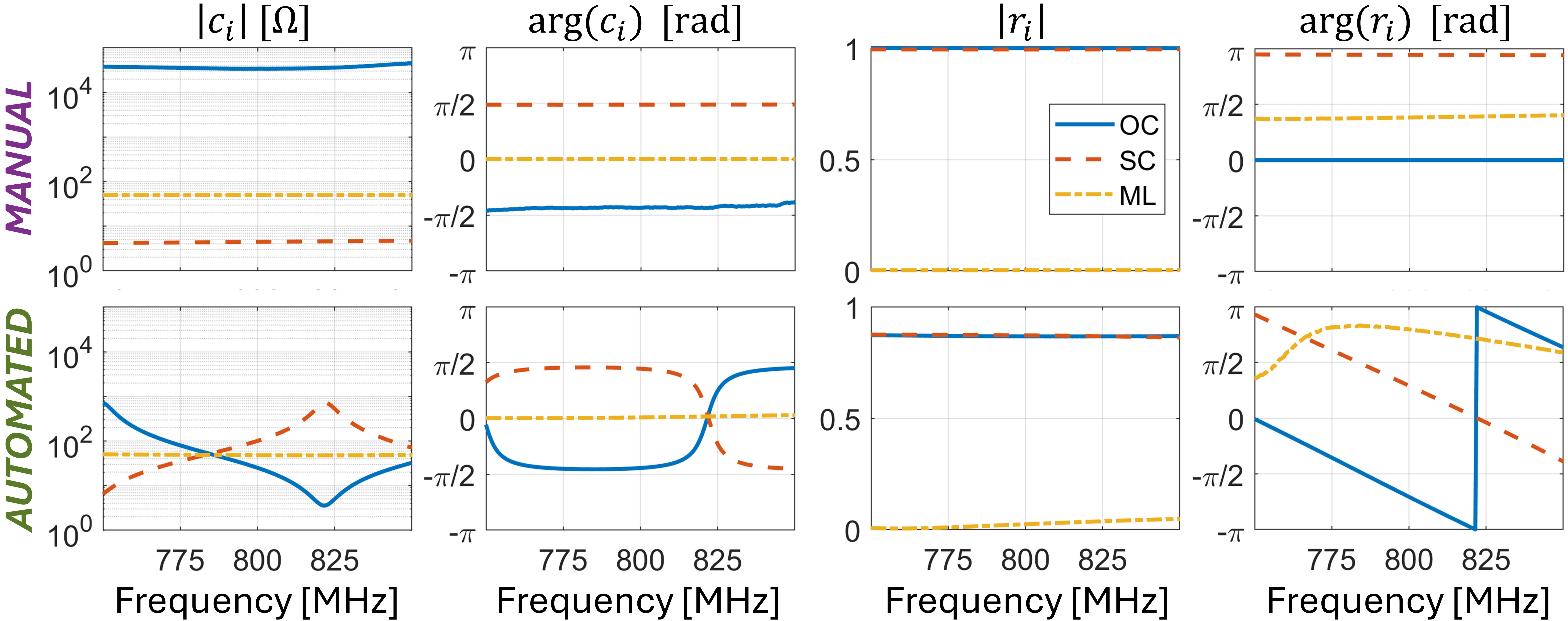}
    \caption{ Magnitude and phase of the impedance $c_i$ and reflection coefficient $r_i$ of the loads seen in Fig.~\ref{Fig1_1}C. The top and bottom row corresponds to manual and automated switching, respectively.}
    \label{Fig1_2}
\end{figure}

We experimentally validate our methods with an eight-port DUT consisting of eight antennas (AEACBK081014-S698) inside a reverberation chamber\footnote{The RC's dimensions are $1.75\mathrm{m}\times 1.50\mathrm{m}\times 2.00\mathrm{m}$ (surface area:~$18.25\mathrm{m}^2$; volume:~$5.25\mathrm{m}^3$). The mode stirrers seen in Fig.~\ref{Fig1_1}A are not used and remain static throughout all experiments.
Based on the decay rate of the average impulse response envelope measured between the antenna pairs, the composite quality factor of the RC is $Q=949$~\cite{west2017best}. We estimate that $n=8\pi V f_0^3 / c^3 Q \approx 3$ modes overlap at any given frequency based on Weyl's law, where $V$, $f_0$ and $c$ denote the RC's volume, the central frequency of the considered interval (i.e., $f_0=800\ \mathrm{MHz}$) and the speed of light, respectively.}, as depicted in Fig.~\ref{Fig1_1}A. Each antenna has a monomodal lumped port with SubMiniature version A (SMA) connector.\footnote{To prevent confusion, we note that the fact that the antenna port is monomodal does not imply any assumptions about the antenna's structural scattering~\cite{king1949measurement,hansen1989relationships,hansen1990antenna}.} The considered DUT is furthermore linear, passive, time-invariant and reciprocal. Hence, all basic requirements for our approach are satisfied. This setup is ideally suited for our proof-of-principle experiments: First, given the chamber's geometric irregularity, it is clear that no DUT-specific a priori knowledge can be used; second, through-wall SMA connectors (see Fig.~\ref{Fig1_1}A,C) enable us to connect the DUT ports to different loads or measurement equipment without perturbing the DUT. We split the DUT's ports into four accessible and four NDA ports. Because we can physically access the NDA ports, we can also connect them directly to our eight-port VNA (Keysight M9005A) and measure the DUT's $8\times 8$ ground truth scattering (or impedance) matrix to validate our results.

For our main experiment, we can terminate each of the DUT's four NDA ports either manually with one of three calibration standards (OC, short circuit, matched load) as seen on the right in Fig.~\ref{Fig1_1}C, or with an electronically tunable one-port circuit built with two electro-mechanical relay switches (PE71S6436) and three calibration standards as seen on the left in Fig.~\ref{Fig1_1}C. The measured reflection and impedance spectra for the three possible loads in the two cases are shown in Fig.~\ref{Fig1_2}. The characteristics of the electronically switchable loads can be seen to clearly \textit{not} emulate calibration standards within the considered frequency range (due to wave propagation in the cables and switches).

Our measurement methodology consists in terminating the four NDA ports with a set of different load configurations (depending on the considered approach, see illustration in Fig.~\ref{Fig0}B), either based on manual or automated switching, and measuring the $4\times 4$ scattering matrix at the DUT's four available ports for each considered load configuration with our VNA (IFBW:~500~MHz; power:~13~dBm). These VNA measurements are exactly the measurements required by our closed-form and gradient-descent methods for complex-valued measurements. Meanwhile, the intensity-only gradient-descent method assumes that the outgoing intensity at the accessible ports is detected upon injecting different known random coherent wavefronts into the accessible ports for a given load configuration. In practice, such a setup could be implemented, for instance, with an ensemble of coherent sources, circulators and a multi-channel spectrum analyzer. For simplicity, however, we exploit the linearity of the system to emulate such measurements in software based on the $4 \times 4$ scattering matrices measured with our VNA.

To the best of our knowledge, no prior experiments on related topics (``unterminating'', ``port reduction'', etc.) have considered a DUT of a comparable complexity and with that many NDA ports.

\subsection{Experimental Results}

We now proceed to compare the three approaches presented in Sec.~\ref{Sec_theory_methods} in terms of the average relative error of the experimentally estimated DUT scattering (or impedance) matrix. All results presented in this section use the technique to lift the ambiguities presented in Sec.~\ref{sssec_directMeas} to facilitate the plotting of phases and the evaluation of errors.

We conduct the comparison of the accuracies achieved with our three approaches as a function of signal-to-noise ratio (SNR), and as a function of the number $N_\mathrm{cal}$ of utilized sets of load configurations in the case of the gradient-descent approaches. (The value of $N_\mathrm{cal}$ is fixed for the closed-form approach.) A visual summary of the utilized sets of load configurations with closed-form and gradient-descent approaches is provided in Fig.~\ref{Fig0}B. We use manual switching for the closed-form approach (such that one of the available loads is a perfect OC, avoiding the need for the extra data analysis step from Sec.~\ref{subsec_gradesc})\footnote{We numerically verified the closed-form approach without availability of an OC load (hence using the additional step from Sec.~\ref{subsec_gradesc}); a recent experimental validation can be found in our follow-up work~\cite{del2024virtual2p0}.}, and we use electronic switching for the gradient-descent approaches.

\begin{figure*}[t]
    \centering
    \includegraphics[width=\textwidth]{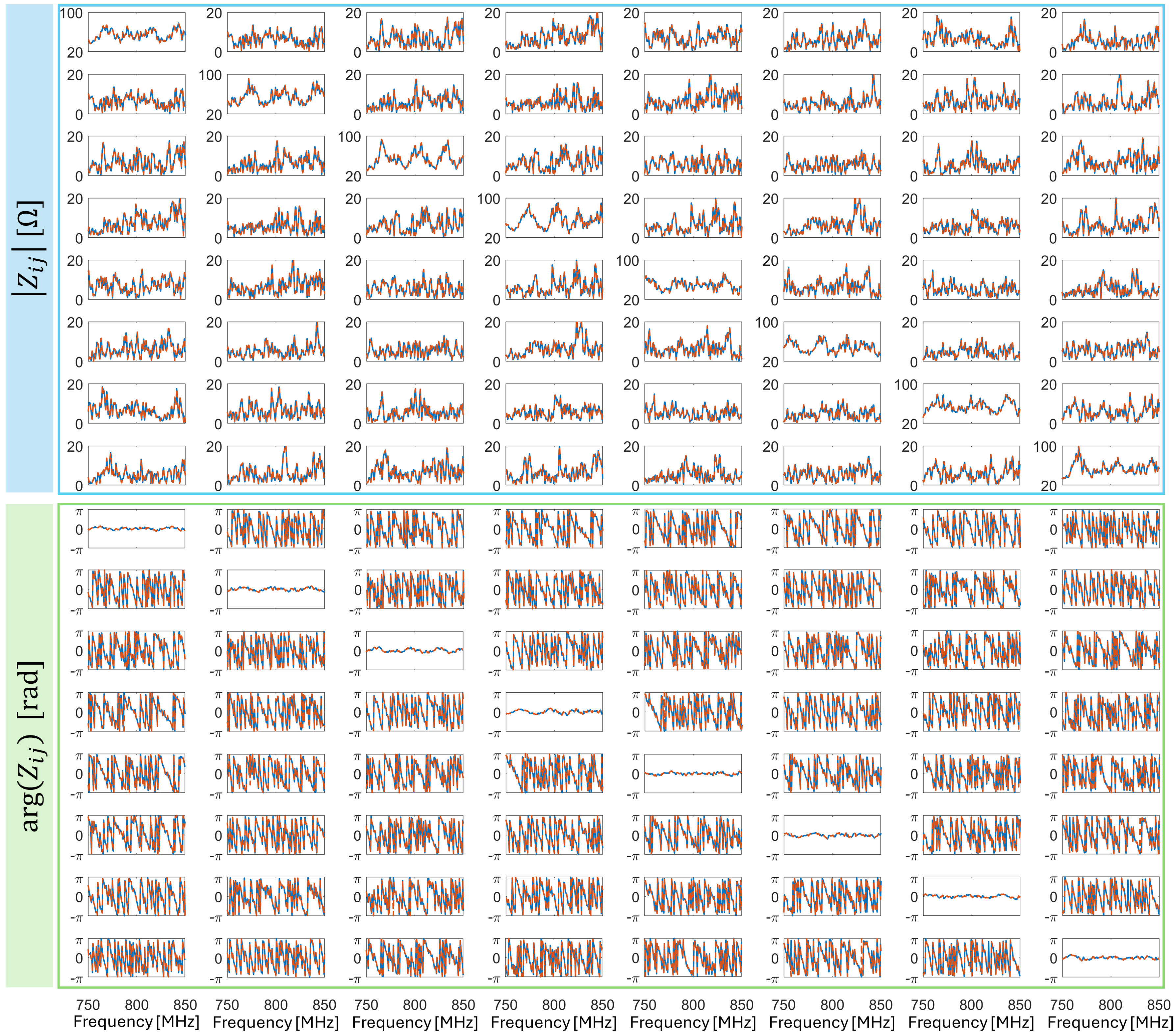}
    \caption{Estimated (red dashed) vs ground-truth (blue continuous) impedance parameter spectra for the setup from Fig.~\ref{Fig1_1}A. The estimated parameters are obtained using the manual load switching (right part of Fig.~\ref{Fig1_1}C) and Algorithm~\ref{Alg1}. The $(i,j)$th subfigure in the top [bottom] panel shows the magnitude [phase] of the $(i,j)$th entry of $\mathbf{Z}$ (denoted by $Z_{ij}$) as a function of frequency. The ambiguities were mitigated with the technique described in Sec.~\ref{sssec_directMeas}.}
    \label{Fig2}
\end{figure*}

\subsubsection{Visual display of estimated impedance matrix}

\begin{figure}[h]
    \centering
    \includegraphics[width=\columnwidth]{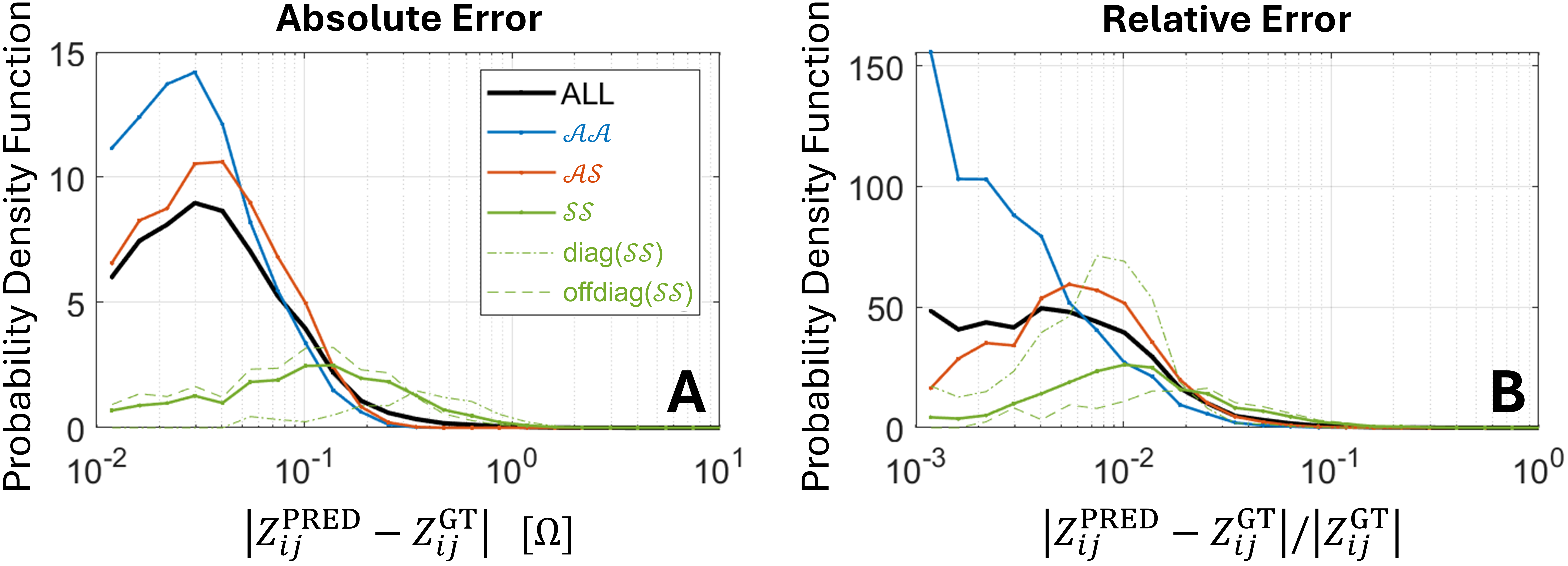}
    \caption{Probability density functions for the absolute (A) and relative (B) errors of the data displayed in Fig.~\ref{Fig2}. The black curves consider all entries of the impedance matrix whereas the colored curves only consider the blocks $\mathcal{AA}$ (blue), $\mathcal{AS}$ (red), or $\mathcal{SS}$ (green). In addition, the green dash-dotted and dashed lines consider only the diagonal or off-diagonal entries of the block $\mathcal{SS}$, respectively.  }
    \label{FigErrDist}
\end{figure}

We begin by displaying in Fig.~\ref{Fig2} the comparison of the estimate of $\mathbf{Z}$ obtained with our closed-form Algorithm~\ref{Alg1} and the ambiguity mitigation from Sec.~\ref{sssec_directMeas}. The corresponding absolute and relative error distributions (including breakdowns by blocks) are displayed in Fig.~\ref{FigErrDist}.
The mean absolute error (averaged over the 64 entries and 201 frequency points) is $0.15\ \Omega$. The blocks $\mathcal{AA}$ and $\mathcal{AS}$ have the lowest average absolute errors of $0.06\ \Omega$ and $0.07\ \Omega$, respectively, while the block $\mathcal{SS}$ has the largest average absolute error of $0.40\ \Omega$. The average absolute error on the diagonal (resp.~off-diagonal) entries of the block $\mathcal{SS}$ is $0.73\ \Omega$ (resp.~$0.29\ \Omega$); yet, the diagonal entries have larger magnitudes than the off-diagonal entries such that the relative errors are in fact lower for the diagonal entries, as seen in Fig.~\ref{FigErrDist}B. 
For compactness, we do not include gradient-descent counterparts to Fig.~\ref{Fig2} based on Algorithm~\ref{Alg2} and Algorithm~\ref{Alg3} here since, upon visual inspection, the reader could not identify any differences compared to Fig.~\ref{Fig2}.

\subsubsection{Systematic analysis}

\begin{figure}[b!]
    \centering
    \includegraphics[width=\columnwidth]{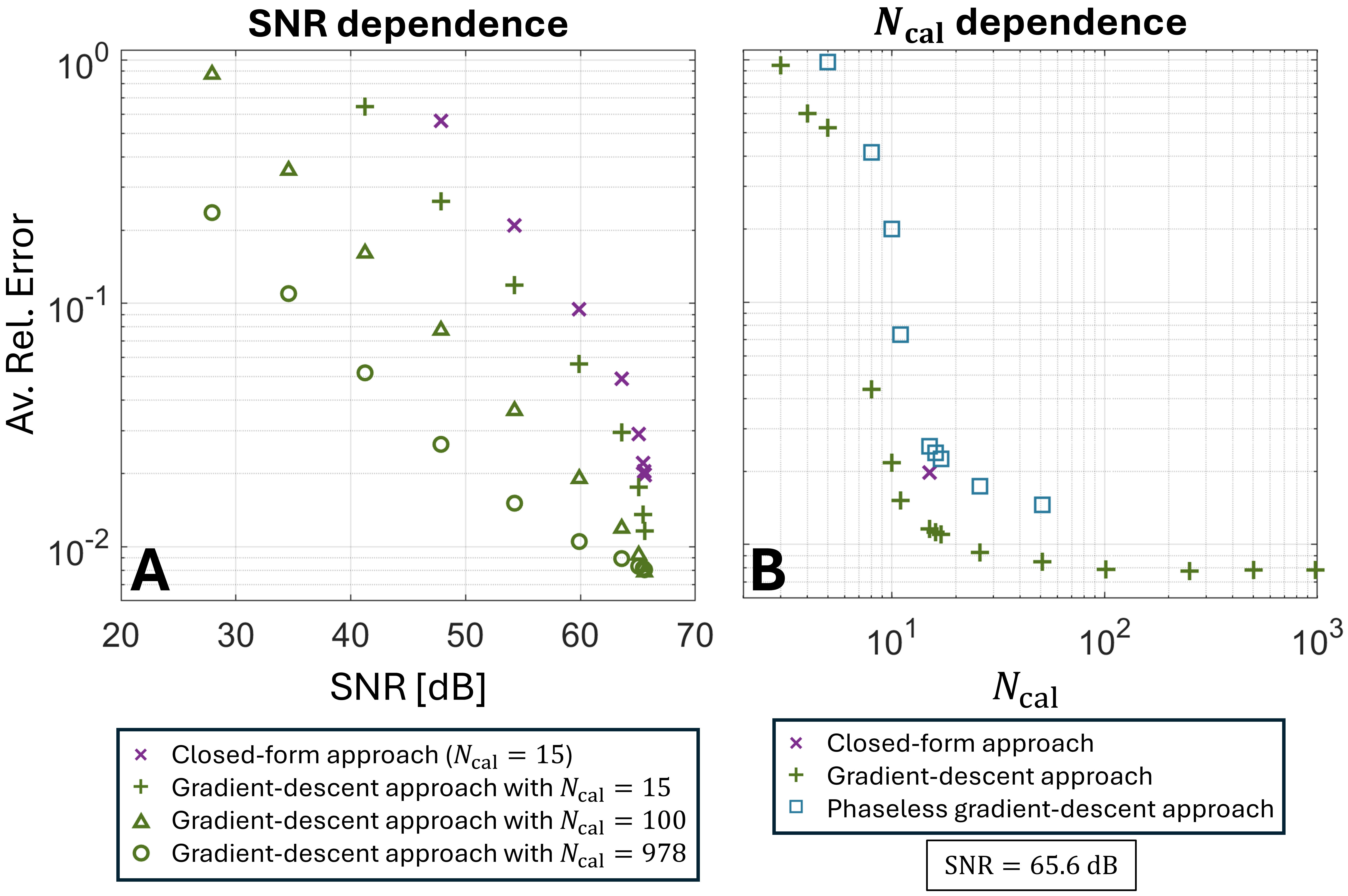}
    \caption{Comparison of closed-form approach (purple,  Algorithm~\ref{Alg1}) vs gradient-descent approach (green,  Algorithm~\ref{Alg2}) vs phaseless gradient-descent approach (blue, Algorithm~\ref{Alg3}, only in (B)) in terms of how the resulting average relative error depends on the SNR (A) and the number of calibration measurements $N_\mathrm{cal}$ (B). Three different values of $N_\mathrm{cal}$ are considered for the gradient-descent approach in (A). The ambiguities were mitigated with the technique described in Sec.~\ref{sssec_directMeas}.}
    \label{Fig3}
\end{figure}

The signal-to-noise ratio (SNR) in our experiments was $65.6\ \mathrm{dB}$. We expect that the accuracy of Algorithm~\ref{Alg1} rapidly deteriorates as the SNR decreases because the magnitude of the changes of $\hat{\mathbf{S}}$ due to the change of one or two load configurations becomes comparable to the noise magnitude, and eventually the noise drowns the sought-after changes of $\hat{\mathbf{S}}$. By manually adding additional measurement noise to the experimental data in post-processing, this expected trend is indeed observed in Fig.~\ref{Fig3}A. Note that we do not add noise to the values of $\mathbf{c}$ because the characterization measurements of the utilized loads can be performed offline under ideal conditions.

While we expect the qualitatively observed SNR dependence to be generic, its quantitative details depend on the considered system. In particular, if the $i$th NDA port is only weakly coupled to the directly accessible ports, then the norm of $\Delta\hat{\mathbf{Z}}_i^{\mathrm{AB}}$ will be particularly low such that the estimation of the latter is particularly sensitive to noise. Specifically, $|\beta_i|$ dictates the coupling strength between the $i$th NDA port and the accessible ports but, of course, the value of $|\beta_i|$ is not known a priori for an unknown DUT.

The average relative error with the gradient-descent approach from Algorithm~\ref{Alg2} is seen in Fig.~\ref{Fig3}A to be below that of the closed-form approach for all considered SNR values. 
We attribute this difference to the fact that the changes of $\Delta\hat{\mathbf{S}}$ are on average larger for the former. In addition, the difference between manual switching for the closed-form approach and automated electronic switching for the gradient-descent approach may matter, too, as our recent follow-up work suggests~\cite{del2024virtual2p0}. We further observe in Fig.~\ref{Fig3}A  that using larger values of $N_\mathrm{cal}$ in the gradient-descent approach can substantially reduce the average relative error in the lower-SNR regime, highlighting the 
advantage of this method's ability to flexibly adapt the value of $N_\mathrm{cal}$  to mitigate noise.

Next, we systematically study the influence of $N_\mathrm{cal}$ on the reconstruction accuracy of the gradient-descent algorithms in Fig.~\ref{Fig3} for the experimental SNR of 65.6~dB. At $N_\mathrm{cal}=15$, which is the value used by the closed-form algorithm, the average relative error is $0.012$ (gradient-descent approach from Algorithm~\ref{Alg2}) instead of $0.020$ (closed-form approach from Algorithm~\ref{Alg1}).
In other words, the gradient-descent approach from Algorithm~\ref{Alg2} can reach the same accuracy as the closed-form approach from Algorithm~\ref{Alg1} with fewer measurements. This effect is observable in Fig.~\ref{Fig3}B but not dramatic because we have only four NDA ports; we expect this effect to become much more prominent for larger values of $N_\mathrm{S}$ (the results in Ref.~\cite{sol2023experimentally} hint at this, although they did not consider ambiguity removal). 
Moreover, the gradient-descent approach from Algorithm~\ref{Alg2} can flexibly increase $N_\mathrm{cal}$ beyond 15 so that the average relative error can drop to $0.008$ in our experiment.

Remarkably, the performance with the intensity-only gradient-descent Algorithm~\ref{Alg3} as a function of $N_\mathrm{cal}$ is seen in Fig.~\ref{Fig3}B to not be significantly worse than that with phase information, reaching an average relative error of 0.015 in the best considered case (albeit at the expense of more measurements than required for the same accuracy when phase information is used, see Fig.~\ref{Fig3}B). For $N_\mathrm{cal} = 15$, the intensity-only gradient-descent algorithm's average relative error of 0.025 is close to the relative average error of 0.020 with the closed-form approach. Overall, the mathematical constraints are hence sufficiently strong to accurately estimate all scattering parameters up to minimal inevitable ambiguities regarding their phases (sign ambiguity on coefficients associated with NDA ports and blockwise phase ambiguity) even without ever measuring phase.

\section{Discussion}
\label{sec_discussion}

\begin{table*}[h!]
\renewcommand{\arraystretch}{1.5} 
\centering
\caption{Comparison with related measurement techniques in terms of their benefits and shortcomings.}
\begin{tabular}{|p{4.5cm}|p{\dimexpr\textwidth-6cm\relax}|} 
\hline
\textbf{Technique} & \textbf{Benefits and Shortcomings} \\ \hline
Use a commercial $N$-port VNA. &
$+$ High accuracy.\newline
$-$  Not applicable to physically inaccessible (embedded) ports.\newline
$-$ Very expensive.\newline
$-$ Limited scalability to DUTs with many ports.\newline
$-$ No option for working with intensity-only data.
\\ \hline
Use a few-port VNA in combination with a commercial full cross-bar switch matrix. &
$+$ High accuracy. \newline
$-$  Not applicable to physically inaccessible (embedded) ports.\newline
$-$ Very expensive.\newline
$-$ Limited scalability to DUTs with many ports.\newline
$-$ No option for working with intensity-only data.
\\ \hline
Manually connect a few-port VNA to different sets of DUT ports, each time terminating the remaining ports with reference loads.~\cite{tippet1982rigorous,ruttan2008multiport,2023paper} &
$-$  Not applicable to physically inaccessible (embedded) ports.\newline
$-$ Prone to human errors.\newline
$-$ Limited scalability to DUTs with many ports.\newline
$-$ Sensitive to quality of reference loads.\newline
$-$ No option for working with intensity-only data.
\\ \hline
``Unterminate'' a single load-tunable port.~\cite{garbacz1964determination,bauer1974embedding,mayhan1994technique,davidovitz1995reconstruction,pfeiffer2005recursive,pfeiffer2005equivalent,pfeiffer2005characterization,pursula2008backscattering,bories2010small,van2020verification,sahin2021noncontact,kruglov2023contactless} &
$-$ Limited to a single NDA port without possibility of dealing with multiple NDA ports.\newline
$-$ Frequent reliance on a priori knowledge about DUT and simplifying approximations.\newline
$-$ No matrix-valued closed-form approach to optimally leverage measurements with a multi-port VNA.\newline
$-$ No gradient-descent approach to flexibly adapt the number of measurements to mitigate noise.\newline
$-$ No option for working with intensity-only data.
\\ \hline
Apply existing multi-port backscatter modulation techniques for antenna characterization using a few-port VNA.~\cite{wiesbeck1998wide,monsalve2013multiport} &
$-$ Reliance on a priori knowledge about DUT and simplifying approximations.\newline
$-$ No matrix-valued closed-form approach to optimally leverage measurements with a multi-port VNA.\newline
$-$ No gradient-descent approach to flexibly adapt the number of measurements to mitigate noise.\newline
$-$ No option for working with intensity-only data.
\\ \hline
Apply ``port-reduction'' methods using a few-port VNA.~\cite{lu2000port,lu2003multiport,pfeiffer2005recursive} &
$-$ The set of ports treated as NDA changes between measurements, requiring manual reconnections and implying a limitation to physically accessible ports.\newline
$-$ No matrix-valued closed-form approach to optimally leverage measurements with a multi-port VNA.\newline
$-$ No gradient-descent approach to flexibly adapt the number of measurements to mitigate noise.\newline
$-$ No option for working with intensity-only data.
\\ \hline
Manually connect one port to a generator and measure radiation pattern for two distinct uniform reference terminations of the other ports, and repeat for each port in turn.~\cite{buck2022measuring} &
$-$ Specific to antenna arrays (not applicable to other multi-port systems like circuits).\newline
$-$ The set of ports treated as NDA changes between measurements, requiring manual reconnections and implying a limitation to physically accessible ports.\newline
$-$ Radiation pattern measurements are costly, time-consuming and potentially of limited accuracy (e.g., if relying on a drone).\newline
$-$ No option for working with intensity-only data.
\\ \hline
Use a protocol for optimal focusing on a load-modulated port.~\cite{abboud2013noniterative,ma2014time,zhou2014focusing,ruan2017focusing,ambichl2017focusing,horodynski2020optimal,del2021coherent,yeo2022time,sol2024optimal} &
$+$ Applicable to embedded NDA ports.\newline
$+$ Only two distinct, uncharacterized loads are required.\newline
$-$ Limited to a single NDA port without possibility of dealing with multiple NDA ports.\newline
$-$ Only estimates the $\mathcal{AS}$ block (which is a vector because of the limitation to $N_\mathrm{S}=1$).\newline
$-$ All estimated magnitudes and phases are ambiguous.\newline
$-$ No option for working with intensity-only data.
\\ \hline
Use a protocol for physics-compliant end-to-end RIS-parametrized channel estimation.~\cite{sol2023experimentally} &
$+$ Applicable to embedded NDA ports.\newline
$+$ Only two distinct, uncharacterized loads are required.\newline
$+$ An option for working with intensity-only data exists.\newline
$-$ All estimated magnitudes and phases are ambiguous.\newline
$-$ No closed-form approach exists so far.
\\ \hline
Use the closed-form ``Virtual VNA'' technique. [This Work] &
$+$ Applicable to embedded NDA ports.\newline
$+$ No requirements for specific loads (but three distinct and known loads are needed).\newline
$+$ Matrix-valued nature allows for optimal exploitation of available number of VNA ports.\newline
$-$ Remaining inevitable sign-ambiguity (can be lifted if problematic, as discussed in Sec.~\ref{sec_discuss_ambiguity}).\newline
$-$ The number of measurements is inflexible and cannot be adapted to mitigate noise but it provides an upper bound on the number of required measurements under ideal condition.\newline
$-$ At most two terminations of NDA ports are different in subsequent measurements, resulting in vulnerability to measurement noise.\newline
$-$ No option for working with intensity-only data.
\\ \hline
Use the gradient-descent ``Virtual VNA'' technique. [This Work] &
$+$ Applicable to embedded NDA ports.\newline
$+$ No requirements for specific loads (but three distinct and known loads are needed).\newline
$+$ The number of measurements is flexible and can be adapted to mitigate noise.\newline
$+$ Compatible with opportunistic load configuration switching.\newline
$+$ Significantly more than two terminations of NDA ports are different in subsequent measurements, resulting in more robustness to measurement noise.\newline
$+$ An option for working with intensity-only data exists (associated with an additional inevitable blockwise phase ambiguity that can be lifted if problematic, as discussed in Sec.~\ref{sec_discuss_ambiguity}).\newline
$-$ Remaining inevitable sign-ambiguity (can be lifted if problematic, as discussed in Sec.~\ref{sec_discuss_ambiguity}).
\\ \hline
\end{tabular}
\label{tab:comparison}
\end{table*}

\begin{figure*}
    \centering
    \includegraphics[width=0.9\textwidth]{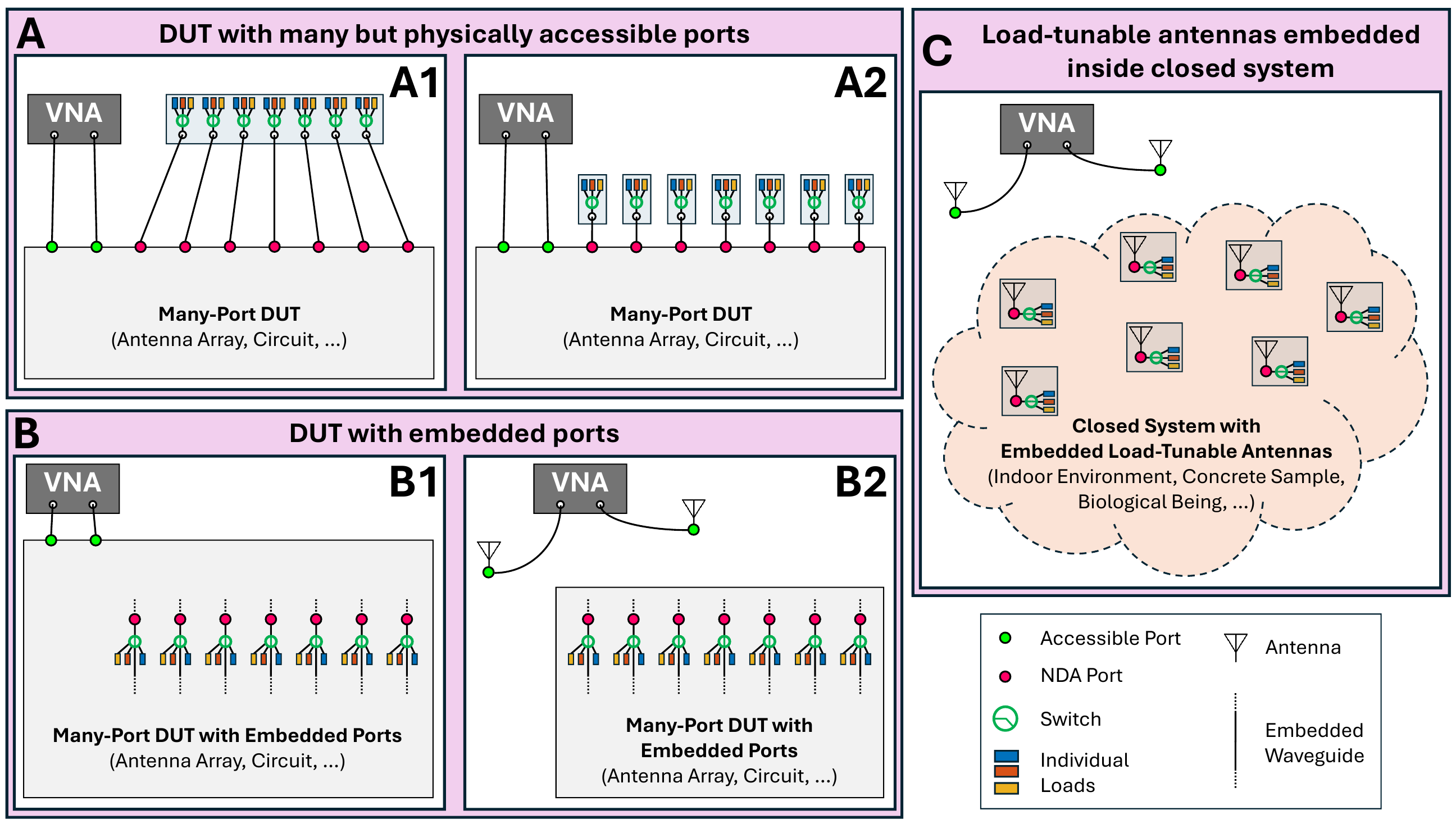}
    \caption{Applications perspectives for the ``Virtual VNA'' techniques. (A) Characterization of a DUT (antenna array, circuit, ...) with many physically accessible ports. A large fraction of the DUT ports is treated as NDA. (A1) The DUT's NDA ports are connected via coaxial cables to a Virtual VNA Extension PCB consisting of a set of ports connected to individually tunable loads with wired powering and control. (A2) Individual untethered tunable loads are attached to each of the DUT's NDA ports. (B) Characterization of a DUT (antenna array, circuit, ...) with physically inaccessible (embedded) ports. The NDA ports are situated on an embedded waveguide and the DUT is engineered such that the NDA ports can be terminated by tunable loads. Wired powering and control of the switching is feasible. (B1) The DUT has a few accessible ports that are directly connected to the VNA. (B2) The DUT has no accessible ports. A few auxiliary antennas directly connected to the VNA are used as accessible ports. (C) Untethered load-tunable antennas are embedded inside a closed system (indoor environment, concrete sample, biological being, ...). The embedded antennas' ports are inaccessible. A few auxiliary antennas directly connected to the VNA are used as accessible ports.}
    \label{FigAppl}
\end{figure*}

A comprehensive comparison of the presented ``Virtual VNA'' techniques to all  related existing techniques mentioned in Sec.~\ref{sec_introduction} and Sec.~\ref{Sec_detailedContextualization} in terms of their benefits and shortcomings is summarized in Table~\ref{tab:comparison}.  
Embodiments of the ``Virtual VNA'' can differ regarding the following two practical aspects:
\begin{enumerate}
    \item \textit{Nature of the NDA Ports:}
    \begin{itemize}
    \item \textit{NDA ports are physically inaccessible.} The NDA ports are physically inaccessible when they are embedded, e.g., inside a printed-circuit board (PCB), a sample (e.g., concrete) or a biological being (i.e., an animal or human subject). In these cases, the tunable loads must be integrated with the port. The loads can be characterized via TRL calibration~\cite{engen1979thru} which requires auxiliary calibration structures. 
    \item \textit{NDA ports are physically accessible.} Ports can be physically accessible but effectively NDA for practical purposes, notably when the DUT has a very large number of ports. In these cases, the tunable loads can be integrated with the port (which is only possible if one engineers the DUT) or connected to the NDA ports (e.g., via SMA connectors as in our experiments shown in Fig.~\ref{Fig1_1}C). The loads can easily be characterized with a one-port VNA.
    \end{itemize}
    \item \textit{Mechanism for Powering and Controlling the Load Configuration Switching:}
\begin{itemize}
    \item \textit{Wired powering and control (tethered).} The mechanism by which the load configuration is switched is both powered and controlled in a tethered manner relying on wires. Our automatically switched experiment shown on the left side in Fig.~\ref{Fig1_1}C falls into this category because we used wires to power the electro-mechanical switches according to the desired configuration. Wired powering and control is conceivable when NDA ports are physically accessible or when NDA ports are physically inaccessible but the system in which they are embedded is engineered (e.g., a PCB, see Fig.~\ref{FigAppl}B). 
    \item \textit{Wireless powering and control (untethered).} By endowing the switching mechanisms with a wireless receiver and rectifier, they can be both powered and controlled wirelessly. The optimal wavefront to wirelessly transfer power from a set of antennas to a given NDA port can be determined based on a single switch of the load terminating that NDA port~\cite{sol2024optimal}. The wireless powering and control of the switches can be realized in between measurements within the same frequency range as the characterization of the DUT.
    \item \textit{Autonomous operation (untethered).} To avoid the need for wired or wireless powering and control, one can endow the tunable loads with a battery and let them cycle through a predefined sequence of configurations~\cite{monsalve2013multiport}. This untethered solution  appears simpler to implement than wireless powering and control but it is less flexible in terms of the utilized configurations and it is limited by the battery lifetime.\end{itemize}
\end{enumerate}

The envisioned application perspectives of the ``Virtual VNA'' techniques are schematically depicted in Fig.~\ref{FigAppl} and summarized as follows:
\begin{enumerate}[label=\Alph*)]
    \item \textit{DUTs with many but physically accessible ports.} A large subset of the DUT's ports are treated as NDA. On the one hand (A1), all required tunable loads can be integrated into a Virtual VNA Extension PCB that is connected via coaxial cables to the DUT's NDA ports. Powering and controlling of the switching can be tethered. On the other hand (A2), each tunable load can be on a separate untethered PCB that is attached directly to an NDA port via a screw connector (e.g., SMA). This can be convenient when the distances between the NDA ports are large.  
    \item \textit{DUTs with embedded ports.} The DUT's NDA ports are situated on integrated waveguides and physically inaccessible. The DUT is engineered such that the NDA ports can be terminated with a tunable load~\cite{kruglov2023contactless}. Powering and controlling of the switching can be tethered. The DUT can either contain a few accessible ports (B1) or a few auxiliary antennas can be used as accessible ports (B2).
    \item \textit{Untethered load-tunable antennas embedded inside a closed system.} A multitude of untethered load-tunable antennas is embedded inside a closed system for sensing based on the embedded antennas' scattering parameters. The closed system could be, for instance, an indoor environment, a sample of concrete, or a biological being. A few auxiliary antennas outside the closed system serve as accessible ports.    
\end{enumerate}
With regards to the ``Virtual VNA 2.0'' concept covered in our recent follow-up work~\cite{del2024virtual2p0} which requires additional coupled loads, we note that cases A1, B1 and B2 from Fig.~\ref{FigAppl} can be modified to include coupled loads.

\section{Conclusion}
\label{sec_conclusion}

To summarize, we have proposed three approaches to estimate with minimal ambiguity the $N\times N$ scattering (or impedance) matrix of a DUT for scenarios in which waves can only be input/output via a fixed set of $N_\mathrm{A}<N$ accessible DUT ports while individual loads terminating the remaining $N_\mathrm{S} = N - N_\mathrm{A}$ NDA DUT ports can be tuned to three distinct and known values. Our only assumptions were that the DUT is linear, passive, time-invariant and reciprocal, and that the DUT's ports are lumped and monomodal. None of our approaches requires any of the three available loads to emulate a special calibration standard, nor that they be identical for each NDA port.

\textit{First,} we established a closed-form approach whose matrix-valued nature can optimally incorporate measurements with arbitrarily many accessible ports, which provides robustness against measurement noise. Using at least two ports also reduces the number of required measurements by $N_S(N_S-1)/2$ to $1+2N_\mathrm{S}+N_S(N_S-1)/2$. The closed-form approach thus establishes an upper bound for the number of required measurements in the low-noise regime, and it clarifies the requirement for at least three distinct and known loads as well as the origin of inevitable remaining sign ambiguities within the considered problem statement. 

\textit{Second,} we established a gradient-descent approach that leverages random (potentially opportunistic) load configurations. This approach appears particularly advantageous to further mitigate measurement noise because the changes of $\hat{\mathbf{S}}$ between subsequent configurations are larger (because typically more than one or two loads were switched), and because it can flexibly incorporate measurements with additional load configurations. In the low-noise regime, this approach can also achieve comparable accuracies to the closed-form approach but using fewer measurements.

\textit{Third,} we established a modified intensity-only gradient-descent approach that dispenses with phase-sensitive measurements, thereby drastically alleviating the detection hardware cost, at the expense of a mild additional blockwise phase ambiguity.

We thoroughly discussed in what applications the remaining inevitable ambiguities are (un)problematic, and different techniques by which they can be mitigated, if needed.
Finally, we experimentally validated our three approaches considering an eight-port DUT based on a reverberation chamber with four accessible and four NDA ports.

Looking forward, \textit{on the algorithmic side}, we expect that a more compact representation of the broadband scattering (or impedance) matrix with a matrix-valued pole-residue model~\cite{Deschrijver2008Macromodeling} will enable us to significantly reduce the number of parameters to be estimated. Incidentally, although this representation would not lift the sign ambiguity, it would ensure that the sign ambiguity is the same at all frequencies.\footnote{Alternatively, with sufficiently fine frequency sampling, such an alignment of the sign ambiguity at all frequencies can also be achieved via the continuity of the phase as a function of frequency~\cite{yeo2022time}.} Moreover, we expect the development of variants of the approaches we presented in this paper for scenarios in which (i) only off-diagonal blocks of $\hat{\mathbf{S}}(\mathbf{r})$ can be measured (such as a wireless channel matrix between a set of transmitting and receiving antennas inside a smart radio environment), (ii) the DUT is non-reciprocal, or (iii) the DUT's ports are multimodal. Meanwhile, \textit{on the hardware side}, we expect that the characteristics of the three utilized loads can be optimized to minimize the sensitivity to noise. For the wireless sensing applications in indoor surveillance, non-destructive testing and wireless bioelectronics described in Sec.~\ref{sec_discuss_ambiguity}, we anticipate the realization of wirelessly programmable and chargeable untethered antennas with three distinct load states. Alternatively, untethered antennas terminated by a tunable load that autonomously cyles through a pre-defined set of configurations are conceivable~\cite{monsalve2013multiport}. In metrology applications concerned with crosstalk characterization, similarly untethered tunable loads could be deployed. In metrology contexts requiring the removal of all ambiguities, we anticipate the realization of the ``Virtual VNA 2.0'' concept covered in our recent follow-up work~\cite{del2024virtual2p0} that additionally includes coupled loads, either as VNA extension kit or directly integrated into next-generation antenna arrays and circuits with embedded ports. Finally, these concepts may proliferate also for other wave phenomena and scales, including the characterization of programmable photonic integrated circuits and optical neural networks, as well as non-invasive optical sensing deep inside complex media.

\appendices

\section{Derivation of Eq.~(\ref{eq__6})}
\label{Appendix_eq6}

To determine $\zeta_i$, we begin by rewriting Eq.~(\ref{eq5}) as follows:
\begin{equation}
    -\beta_i^{-2} = \left(k_i^\mathrm{AB}\right)^{-1} \left[\left(\zeta_i + c_i^\mathrm{A} \right)^{-1} - \left(\zeta_i + c_i^\mathrm{B} \right)^{-1} \right]. 
    \label{eq_S4}
\end{equation}
While the left hand side of Eq.~(\ref{eq_S4}) is independent of A and B, we can have different realizations of the right hand side by choosing different pairs of realizations (i.e., \{A,B\}, \{A,C\}, \{B,C\}). Without loss of generality, we chose the two pairs \{A,B\} and \{A,C\}, yielding
\begin{equation}
\begin{split}
   &\left(k_i^\mathrm{AB}\right)^{-1} \left[\left(\zeta_i + c_i^\mathrm{A} \right)^{-1} - \left(\zeta_i + c_i^\mathrm{B} \right)^{-1} \right] \\= &\left(k_i^\mathrm{AC}\right)^{-1} \left[\left(\zeta_i + c_i^\mathrm{A} \right)^{-1} - \left(\zeta_i + c_i^\mathrm{C} \right)^{-1} \right],
   \end{split}
    \label{eq_S5}
\end{equation}
where the only unknown is $\zeta_i$. Multiplying both sides of Eq.~(\ref{eq_S5}) by $k_i^\mathrm{AB}k^\mathrm{AC}\left(\zeta_i + c_i^\mathrm{A} \right)\left(\zeta_i + c_i^\mathrm{B} \right)\left(\zeta_i + c_i^\mathrm{C} \right)$ yields
\begin{equation}
\begin{split}
   &k_i^\mathrm{AC} \left[\left(\zeta_i + c_i^\mathrm{B} \right)\left(\zeta_i + c_i^\mathrm{C} \right) - \left(\zeta_i + c_i^\mathrm{A} \right)\left(\zeta_i + c_i^\mathrm{C} \right) \right] \\=  &k_i^\mathrm{AB} \left[\left(\zeta_i + c_i^\mathrm{B} \right)\left(\zeta_i + c_i^\mathrm{C} \right) - \left(\zeta_i + c_i^\mathrm{A} \right)\left(\zeta_i + c_i^\mathrm{B} \right) \right].
   \end{split}
    \label{eq_S6}
\end{equation}
Straightforward algebraic manipulations of Eq.~(\ref{eq_S6}) lead to
\begin{equation}
\begin{split}
    &k_i^\mathrm{AC} \left[ \zeta_i\left( c_i^\mathrm{B} - c_i^\mathrm{A} \right) + c_i^\mathrm{C}\left( c_i^\mathrm{B} - c_i^\mathrm{A} \right) \right] \\=   &k_i^\mathrm{AB} \left[ \zeta_i \left( c_i^\mathrm{C} - c_i^\mathrm{A} \right) + c_i^\mathrm{B}\left( c_i^\mathrm{C} - c_i^\mathrm{A} \right) \right]
    \end{split}
    \label{eq_S7}
\end{equation}
and finally to Eq.~(\ref{eq__6a}).

Once the value of $\zeta_i$ is determined via Eq.~(\ref{eq__6a}), using Eq.~(\ref{eq_S4}) straightforwardly yields  Eq.~(\ref{eq__6b}). 
Under ideal noiseless conditions, all realization pairs \{A,B\} should yield the exact same value of $\beta_i^2$. Alternatively, we can define $\beta_i^2$ as the average of the three values obtained with the three possible realization pairs: \{A,B\}, \{A,C\}, \{B,C\}.

\section{Derivation of Eq.~(\ref{eq9})}
\label{Appendix_eq9}

Let $\hat{\mathbf{Z}}_{ij}^\mathrm{BB}\in \mathbb{C}^{N_\mathrm{A} \times N_\mathrm{A}}$ denote the measurable impedance matrix for the configuration in which the $i$th and $j$th NDA ports are terminated by load impedances $c_i^\mathrm{B}\neq c_\mathrm{OC}$ and $c_j^\mathrm{B}\neq c_\mathrm{OC}$, respectively, and all other NDA ports are terminated by OC. Then,
\begin{equation}
\begin{split}
    \hat{\mathbf{Z}}_{ij}^\mathrm{BB} &= {\mathbf{Z}}_\mathcal{AA} - \begin{bmatrix}    \mathbf{z}_i &  \mathbf{z}_j \end{bmatrix} \left( \begin{bmatrix}    \zeta_i & \kappa_{ij} \\ \kappa_{ij} &  \zeta_j     \\ \end{bmatrix} + \begin{bmatrix}    c_i^\mathrm{B} & 0 \\ 0 &  c_j^\mathrm{B}     \\ \end{bmatrix} \right)^{-1} \begin{bmatrix}    \mathbf{z}_i^T \\  {\mathbf{z}}_j^T \\ \end{bmatrix} \\& = {\mathbf{Z}}_\mathcal{AA} - \begin{bmatrix}    \mathbf{z}_i &  \mathbf{z}_j \end{bmatrix}    \left(\chi        \begin{bmatrix}    \zeta_j+c_j^\mathrm{B} & -\kappa_{ij} \\ -\kappa_{ij} &  \zeta_i+c_i^\mathrm{B}    \\ \end{bmatrix} \right)  \begin{bmatrix}    \mathbf{z}_i^T \\  {\mathbf{z}}_j^T \\ \end{bmatrix},
    \label{Zaug_24}
    \end{split}
\end{equation}
where we analytically inverted the $2\times 2$ matrix in the second step and $\chi = \frac{1}{(\zeta_i+c_i^\mathrm{B})(\zeta_j+c_j^\mathrm{B}) - \kappa_{ij}^2}$. The rank-two update $\mathbf{D}$ of the measurable impedance matrix upon switching the $i$th and $j$th NDA ports from OC terminations to $c_i^\mathrm{B}$ and $c_j^\mathrm{B}$, respectively, is hence given by
\begin{equation}
    \mathbf{D} = {\mathbf{Z}}_\mathcal{AA} - \hat{\mathbf{Z}}_{ij}^\mathrm{BB}
    =  \begin{bmatrix}    \mathbf{z}_i &  \mathbf{z}_j \end{bmatrix}    \left(\chi        \begin{bmatrix}    \zeta_j+c_j^\mathrm{B} & -\kappa_{ij} \\ -\kappa_{ij} &  \zeta_i+c_i^\mathrm{B}     \\ \end{bmatrix} \right)  \begin{bmatrix}    \mathbf{z}_i^T \\  \mathbf{z}_j^T \\ \end{bmatrix}.
    \label{eq_S16}
\end{equation}

Next, assuming $N_\mathrm{A}>1$, we evaluate 
\begin{equation}
    \mathbf{Q} =  {\mathbf{Z}}_\mathcal{A\bar{S}}^+  \mathbf{D} {\mathbf{Z}}_\mathcal{\bar{S}A}^+ = \chi       \begin{bmatrix}    \zeta_j+c_j^\mathrm{B} & -\kappa_{ij} \\ -\kappa_{ij} &  \zeta_i+c_i^\mathrm{B}     \\ \end{bmatrix} ,
    \label{eq_S17}
\end{equation}
where ${\mathbf{Z}}_\mathcal{A\bar{S}}^+$ denotes the Moore–Penrose pseudo-inverse of ${\mathbf{Z}}_\mathcal{A\bar{S}} = \begin{bmatrix}    \mathbf{z}_i &  \mathbf{z}_j \end{bmatrix} $ and $\mathcal{\bar{S}} = \{i,j\}$. Since ${\mathbf{Z}}_{\mathcal{A\bar{S}}}$ has linearly independent columns, there is an exact expression for ${\mathbf{Z}}_{\mathcal{A}\mathcal{S}_{ij}}^+$:
\begin{equation}
    {\mathbf{Z}}_\mathcal{A\bar{S}}^+ = \left( {\mathbf{Z}}_\mathcal{A\bar{S}}^\dagger {\mathbf{Z}}_\mathcal{A\bar{S}}  \right)^{-1} {\mathbf{Z}}_\mathcal{A\bar{S}}^\dagger \in \mathbb{C}^{2 \times N_\mathrm{A}}.
\end{equation}
Similarly,
\begin{equation}
    {\mathbf{Z}}_\mathcal{\bar{S}A}^+ = {\mathbf{Z}}_\mathcal{\bar{S}A}^\dagger \left( {\mathbf{Z}}_\mathcal{\bar{S}A} {\mathbf{Z}}_\mathcal{\bar{S}A}^\dagger  \right)^{-1} \in \mathbb{C}^{N_\mathrm{A} \times 2}.
\end{equation}

The only unknown in Eq.~(\ref{eq_S17}) is $\kappa_{ij}$. Using the notation $\mathbf{Q} = \begin{bmatrix}    Q_{11} & Q_{12} \\ Q_{21} &  Q_{22}     \\ \end{bmatrix}$, it follows directly from Eq.~(\ref{eq_S17}) that
\begin{subequations}
    \begin{equation}
        Q_{11} = \frac{\zeta_j + c_j^\mathrm{B}}{(\zeta_i+c_i^\mathrm{B})(\zeta_j+c_j^\mathrm{B}) - \kappa_{ij}^2}.
        \label{eq_S20a}
    \end{equation}
    \begin{equation}
        Q_{22} = \frac{\zeta_i + c_i^\mathrm{B}}{(\zeta_i+c_i^\mathrm{B})(\zeta_j+c_j^\mathrm{B}) - \kappa_{ij}^2}.
        \label{eq_S20b}
    \end{equation}
        \begin{equation}
        Q_{12} = Q_{21} = \frac{-\kappa_{ij}}{(\zeta_i+c_i^\mathrm{B})(\zeta_j+c_j^\mathrm{B}) - \kappa_{ij}^2}.
        \label{eq_S20c}
    \end{equation}
\end{subequations}
We can solve Eq.~(\ref{eq_S20a}) and/or  Eq.~(\ref{eq_S20b}) for $\kappa_{ij}^2$:
\begin{equation}
\begin{split}
    \kappa_{ij}^2 &= (\zeta_i+c_i^\mathrm{B})(\zeta_j+c_j^\mathrm{B}) - \frac{\zeta_j + c_j^\mathrm{B}}{Q_{11}} \\&= (\zeta_i+c_i^\mathrm{B})(\zeta_j+c_j^\mathrm{B}) - \frac{\zeta_i + c_i^\mathrm{B}}{Q_{22}}.
    \end{split}
    \label{eq_S21}
\end{equation}
Finally, we solve Eq.~(\ref{eq_S20c}) for $\kappa_{ij}$,
\begin{equation}
    \kappa_{ij} = -Q_{12} \left( (\zeta_i+c_i^\mathrm{B})(\zeta_j+c_j^\mathrm{B}) - \kappa_{ij}^2  \right),
    \label{eq_S22}
\end{equation}
and insert the value for $\kappa_{ij}^2$ obtained in Eq.~(\ref{eq_S21}) into Eq.~(\ref{eq_S22}). Since we did not introduce $\mathbf{Q}$ in the main text, let us explicitly point out that Eq.~(\ref{eq_S22}) and Eq.~(\ref{eq9}) are identical; moreover, Eq.~(\ref{eq_S21}) and Eq.~(\ref{eq10}) are identical.

\section{Gradient descent algorithm}
\label{appendixB}

We use the TensorFlow library to implement the gradient descent with an error back-propagation algorithm. Each frequency point is treated separately. Our cost function to be minimized for a given frequency point is defined as
\begin{equation}
    C = \frac{\left\langle \left| \Delta\hat{\mathbf{S}}^\mathrm{AB}_\mathrm{pred} - \Delta\hat{\mathbf{S}}^\mathrm{AB}_\mathrm{true}  \right| \right\rangle_\mathrm{AB} }{ \left\langle \left| \Delta\hat{\mathbf{S}}^\mathrm{AB}_\mathrm{true}  \right| \right\rangle_\mathrm{AB}},
    \label{eq13}
\end{equation}
where $\Delta\hat{\mathbf{S}}^\mathrm{AB}_\mathrm{true}$ and $\Delta\hat{\mathbf{S}}^\mathrm{AB}_\mathrm{pred}$ are the ground truth and model prediction for $\Delta\hat{\mathbf{S}}^\mathrm{AB}$ [see Eq.~(\ref{eq11})]. The average is taken over the $N_\mathrm{cal}-1$ available realizations of triplets $\{\mathbf{r}^\mathrm{A},\mathbf{r}^\mathrm{B},\Delta\hat{\mathbf{S}}^\mathrm{AB}\}$. We initialize all variables of the model (i.e., all unknown parameters) randomly with values from a truncated normal distribution (mean:~0;~standard deviation:~0.1). The only constraint we impose is reciprocity. For training, we use a batch size of $\mathrm{min}(N_\mathrm{cal}-1,100)$ and the Adam method for stochastic optimization with a step size of $10^{-3}$. We use all available $N_\mathrm{cal}-1$ realizations as training data and stop training after 4000 iterations. For $N_\mathrm{cal}=15$, the estimation of all $N_\mathrm{params} = N(N-1) = 56$ parameters (counting each complex-valued unknown as two real-valued parameters to be estimated) takes about six seconds on a desktop computer with an AMD Ryzen 7 3700x processor and 16 GB RAM.

\section{Details on phase retrieval}
\label{appendixC}

We use the TensorFlow library to implement the gradient descent necessary to identify the phases of $\hat{\mathbf{S}}^\mathrm{A}$ up to a global phase uncertainty $\theta^\mathrm{A}$ based on phase-insensitive measurements. Our cost function is the magnitude of the difference between the predicted and measured intensities of the outputs $\hat{\mathbf{y}}\in\mathbb{C}^{N_\mathrm{A}\times 1}$, averaged over 100 random but known inputs $\hat{\mathbf{x}}\in\mathbb{C}^{N_\mathrm{A}\times 1}$, where $\hat{\mathbf{y}} = \hat{\mathbf{S}}^\mathrm{A}\hat{\mathbf{x}}$. (The choice to use 100 inputs is somewhat arbitrary; we did not seek to minimize this number in this work since our goal here is merely to show the feasibility of phase-insensitive parameter estimation in the considered problem.) The only constraint we impose is reciprocity. We train with a batch size of 100 for 4000 iterations using the Adam method for stochastic optimization with a step size of $10^{-3}$.

To ensure that all realizations of $\hat{\mathbf{S}}^\mathrm{A}$ have the same global phase ambiguity $\theta^\mathrm{A}$, we consider quadruples $\{\hat{\mathbf{S}}^\mathrm{A},\mathbf{r}^\mathrm{A},\hat{\mathbf{S}}^\mathrm{B},\mathbf{r}^\mathrm{B}\}$ for which $\mathbf{r}^\mathrm{A}$ and $\mathbf{r}^\mathrm{B}$ differ by at least one and at most $N_\mathrm{S}-1$ entries. Knowing that the rank $k$ of $\Delta\mathbf{S}^\mathrm{AB}$ is equal to the number of entries by which $\mathbf{r}^\mathrm{A}$ and $\mathbf{r}^\mathrm{B}$ differ, we sweep through all possible values of $\theta^\mathrm{B}$ and pick the one for which the ratio between the $k$th and $(k+1)$th singular values of $\Delta\mathbf{S}^\mathrm{AB}$ is maximal.\footnote{An alternative approach not requiring knowledge of $k$ consists in choosing the value of $\theta^\mathrm{B}$ that minimizes the effective rank~\cite{roy2007effective} of $\Delta\mathbf{S}^\mathrm{AB}$.} Without loss of generality, we pick the first realization as reference configuration and align the global phase offsets of all other realizations with the first one. If for some realization $\mathbf{r}^\mathrm{B}$ differs in all $N_\mathrm{S}$ entries from $\mathbf{r}^\mathrm{A}$, we use a different previously aligned realizations for which $1<k<N_\mathrm{S}$.

Since the global phase $\theta$ of the retrieved complex-valued data is now fixed but unknown and arbitrary, we update the cost function from Eq.~(\ref{eq13}) to be insensitive to $\theta$. This alleviates the difficulty of the gradient descent optimization. We define the following modified cost function
\begin{equation}
    C^\mathrm{PL} = \frac{\left\langle \left| \mathrm{e}^{\jmath\gamma^\mathrm{AB}}\Delta\hat{\mathbf{S}}^\mathrm{AB}_\mathrm{pred} - \Delta\hat{\mathbf{S}}^\mathrm{AB}_\mathrm{true}  \right| \right\rangle_\mathrm{AB} }{ \left\langle \left| \Delta\hat{\mathbf{S}}^\mathrm{AB}_\mathrm{true}  \right| \right\rangle_\mathrm{AB}},
    \label{eq14}
\end{equation}
where
\begin{equation}
    \gamma^\mathrm{AB} = \mathrm{arg} \left(\mathrm{Tr}\left( \left(\Delta\hat{\mathbf{S}}^\mathrm{AB}_\mathrm{pred}\right)^* \Delta\hat{\mathbf{S}}^\mathrm{AB}_\mathrm{true} \right)\right).
    \label{eq15}
\end{equation}
Using the cost function $C^\mathrm{PL}$ in the main gradient descent part implies blockwise phase ambiguities of $\theta$ and $\phi/2$ on $\mathbf{S}_\mathcal{SS}$ and $\mathbf{S}_\mathcal{AS} = \mathbf{S}_\mathcal{SA}^T$, respectively, where generally $\theta \neq \phi$. Fortunately, we can simply sweep through all possible values of $\theta$ and retain the one for which the $N_\mathrm{cal}$ values estimated for $\mathbf{S}_\mathcal{AA}$ with Eq.~(\ref{eq12}) are the most similar to each other (ideally, they are all identical). Thereby, we ensure that the global phase offset of the $\mathbf{S}_\mathcal{AA}$ block is $\phi$. Then, the situation is as described in the main text for $\theta$:  we expect a blockwise phase offset of $\phi$ on $\mathbf{S}_\mathcal{AA}$ because of the blockwise phase offset of $\phi/2$ on $\mathbf{S}_\mathcal{AS}=\mathbf{S}_\mathcal{SA}^T$.

\section*{Acknowledgment}
The author acknowledges stimulating discussions with S.~Bories, H.~Prod'homme and K.~Warnick.

\bibliographystyle{IEEEtran}

\providecommand{\noopsort}[1]{}\providecommand{\singleletter}[1]{#1}%

\end{document}